\documentclass[12pt]{article}
\usepackage{epsfig,dina4}

\begin{document}

\title{\Large 
Extensive Air Shower Simulations at the Highest Energies}

\author{ 
J. Knapp$^1$\thanks{Corresponding author: j.knapp@leeds.ac.uk}, 
D. Heck$^2$, 
S.J. Sciutto$^3$, 
M.T. Dova$^3$ and 
M. Risse$^2$ \\[3mm]
\small $^1$Dept. of Physics and Astronomy, University of Leeds, 
Leeds LS2 9JT, United Kingdom\\
\small $^2$ Institut f\"ur Kernphysik, Forschungszentrum Karlsruhe,
D-76021 Karlsruhe, Germany\\
\small $^3$  Dept. de Fisica, Universidad Nacional de La Plata, 
C.C. 67 - 1900 La Plata, Argentina
}
\date{}
\maketitle
\begin{abstract}
Air shower simulation programs are essential tools for the analysis of
data from cosmic ray experiments and for planning the layout of new detectors.
They are used to estimate the energy and mass of the primary particle.
Unfortunately the model uncertainties
translate directly into systematic errors in the energy and mass
determination.  Aiming at energies $> 10^{19}$ eV, the models have to be
extrapolated far beyond the energies available at accelerators.  On the
other hand, hybrid measurement of ground particle densities and
calorimetric shower energy, as will be provided by the Pierre Auger
Observatory, will strongly constrain shower models.  While the main
uncertainty of contemporary models comes from our poor knowledge of the
(soft) hadronic interactions at high energies, also electromagnetic
interactions, low-energy hadronic interactions and the particle
transport influence details of the shower development.  We review here
the physics processes and some of the computational techniques of
air shower models presently used for highest energies, and discuss the
properties and limitations of the models.
\end{abstract}


\maketitle

\section{Introduction}
The cosmic ray (CR) energy spectrum extends up to $3\times 10^{20}$
eV. The presence of the highest-energy particles (Ultra High-Energy CRs,
UHECRs) poses an enigma, since many good arguments suggest that they should
not be observed.  This apparent contradiction has stimulated a variety 
of more exotic explanations of their existence. 
The enigma can only be solved by an
experiment that can provide a much larger event statistics than the
about 20 events with $E > 10^{20}$ eV measured in the past 35
years. Knowing the form of the energy spectrum of the CR particles,
their arrival direction distribution over the whole sky, and possibly
even their mass composition, would allow us to test some of the
hypotheses about their origin and help to identify their sources. At
present experimental results suggest that UHECRs are protons or nuclei,
as for CRs at much lower energies.  However, many of the more exotic
models of UHECR origin predict also photons and neutrinos.

The measurement of extensive air showers (EAS) is presently the only way
to study CRs with energies above about 10$^{15}$ eV. The properties of
primary cosmic rays have to be deduced from the development of the 
shower in the atmosphere and from the particle ratios in the shower.
The incident direction can easily be reconstructed
from the arrival times of shower particles at different positions at the
observation level and the primary energy is approximately reflected in
the total number of secondary particles produced.  The mass of the
primary particle is more difficult to measure. It is reflected, in a subtle
way, in the shower form, specifically the height of the shower maximum,
and in the muon-to-electron ratio of the shower.

The Pierre Auger Observatory is conceived to measure CRs with energies
$> 10^{19}$ eV with good statistics over the whole sky \cite{auger}. It
will consist of two detector sites, one in the southern and one in the
northern hemisphere.  Each site covers an area of 3000 km$^2$ and
combines two techniques to measure (i) the particle distribution at
observation level with an array of water Cherenkov detectors and (ii)
the longitudinal shower development via optical imaging of the
fluorescence light in the atmosphere during clear moonless nights ($\approx$
10\% of the total time). This hybrid detection provides a way to
inter-calibrate both sub-systems and to control systematic
uncertainties.  The energy determination via the fluorescence light is
basically calorimetric and therefore much less model-dependent than the
energy reconstruction from particle densities at ground level.  The
southern site is presently under construction in Argentina. The size of
the site was chosen to register about 5000 events with energies $>
10^{19}$ eV and about 40-80 events above 10$^{20}$ eV per year.

Since experiments at energies $> 10^{15}$ eV cannot be calibrated with a
test beam the interpretation of EAS measurements is performed by
comparing experimental data with model predictions of the shower
development in the atmosphere. Therefore quantitative results rely on
the model assumptions and on the quality of the simulation of particle
interactions and transport in the atmosphere.

The detailed shower development is far too complex to be fully described
by a simple analytical model. Therefore it is usually modeled by Monte
Carlo (MC) simulation of transport and interaction of each individual
shower particle, employing our present knowledge on interactions, decays
and particle transport in matter.  While the electromagnetic interaction
(responsible for electromagnetic sub-showers, ionization, Cherenkov
light production, ...) and the weak interaction (responsible for decays
of unstable secondaries) are well understood, the major
uncertainties in EAS simulation arise from the hadronic interaction
models.  With the present theoretical understanding of soft hadronic
interactions, i.e. those with a small momentum transfer, one cannot 
calculate interaction cross-sections or particle
production from first principles. Therefore, hadronic interaction models
are usually a mixture of fundamental theoretical ideas and empirical
parametrisations tuned to describe the experimental data at lower
energies.  The large extrapolation (over 6 orders of magnitude in
energy) needed from experimental accelerator data to CR interactions is
the second major source of uncertainty, and with an uncertain
interaction model it is difficult to determine the energy spectrum and
the composition of CRs.

As a consequence of the better understanding of hadronic and
nuclear interactions at high energies, and the increase in computing
power, shower models have improved dramatically over recent years.  Also the
understanding of the measuring process with a detector has
markedly advanced. It is now possible to describe consistently
experimental results over a wide range of energies and even to test
hadronic models on the 20\% level.

In this paper we review the physics and techniques of state-of-the-art
modeling of extensive air showers, specifically those important for
energies of $10^{19}$ eV or above.  In Sec. \ref{sec-had} and
\ref{sec-em} we describe the modeling of hadronic and electromagnetic
interactions, respectively.  Sec. \ref{sec-neu} summarizes briefly the
simulation of showers induced by photons and neutrinos or other more
exotic primary particles.  Statistical thinning techniques that allow 
simulation of showers at the highest energies in a finite time are
discussed in Sec.  \ref{sec-thin}.  Then two shower simulation programs
are described and compared in Sec. \ref{sec-programs}. Some
simulation results are shown in Sec. \ref{sec-results}.

\section{Hadronic interaction models}
\label{sec-had}

The highest energy reached in a man-made accelerator is at present about
$E_{\rm lab}=900$ GeV.  This is about 8 orders of magnitude smaller than
the highest energy ever measured for a CR particle. Events triggered and
examined at accelerator experiments are those that produce particles
with high momentum transfer. They are well described by QCD but they
constitute only a minute fraction ($\ll 10^{-6}$) of the overall
reaction rate. Interactions with low momentum transfers, i.e. soft
collisions, produce particles with small transverse momenta that mostly
escape undetected in the beam pipe.  Of special importance are the
diffractive dissociation events, which originate from rather peripheral
collisions with a small fraction of energy transferred into secondary
particles.  But these reactions are important for air showers, since
they carry the energy deep down into the atmosphere and thus drive the
air shower development.  Moreover, CRs collisions are predominantly
nucleon-nucleus or nucleus-nucleus collisions, for which accelerator
results are available only at much lower energies, rather than
proton-proton collisions.  Models based on accelerator results have to
be extrapolated far into unknown territory in energy, in the kinematic
range of very forward particle production, and to other
projectile-target combinations.  That is why it is of utmost importance
that models rely on a sound theoretical basis which gives some
guidelines on how the interactions evolve with energy.

Early hadronic models have been purely phenomenological. Accelerator
results were parametrised and crudely extended to deal with diffractive
interactions and nucleon-nucleus collisions. Primary nuclei have been
treated with the assumption that a nucleus is just a superposition of
free nucleons. Then the parametrisations have been extrapolated to high
energies. It was recognised that the development of an EAS for a given
energy and primary is mainly dependent on two factors: on the inelastic
cross-sections $\sigma_{\rm inel}$ of primary and secondary particles
with air and on the average fraction of the available energy transferred
into secondary particles (usually termed {\em inelasticity} $k_{\rm
inel}$). Some of the models even had $\sigma_{\rm inel}$ and $k_{\rm
inel}$ as their main free parameters.  With these simple models
occasionally even very basic experimental data just could not be
explained at all and therefore a proper interpretation of the data was
impossible.

Over the last 10 years this situation has changed. Microscopic models
have been developed that are based more and more on reliable theoretical
foundations and attempt to describe not only CR showers but equally well
heavy ion collision or other particle interactions. These models have in
general a much smaller number of free parameters reducing the
arbitrariness, and allow, to some extent, to predict $\sigma_{\rm inel}$
as a function of energy and to deduce the average value of $k_{\rm
inel}$ as well as its distribution from deeper-lying principles.

The new models simulate interactions of nucleons and nuclei on the basis
of the Gribov-Regge theory (GRT) \cite{gribov}, which most successfully
describes elastic scattering and, via the optical theorem, the total
hadronic cross-section as a function of energy. Tab. 38.2 in the
particle data book \cite{pdg} demonstrates good agreement for all
possible scattering processes examined with the Reggeon-Pomeron
scattering scheme. In GRT the observed rise of the cross-sections at
high energies is a consequence of the exchange of multiple {\em
super-critical Pomerons}. Inelastic processes are described by {\em cut
Pomerons} leading to two colour strings each, which fragment subsequently
to colour neutral hadrons. The probability of n exchanged and m cut
Pomerons is uniquely predicted by the theory.  While this is common for
all implementations of the GRT, it is still debated how to realize best
diffractive events and the production and decay of colour strings.
Presently the Gribov-Regge approach is the only theoretically sound way,
and also the most successful one, to model energetic soft hadronic
interactions.  A good introduction to GRT models is given in ref.
\cite{venus}.  At present GRT-type models used for cosmic rays are
QGSJET \cite{qgsjet,kalm}, VENUS \cite{venus}, DPMJET \cite{dpmjet}, and,
most recently, {\sc neXus} \cite{nexus}.

Present work on GRT models is focused on the consistent treatment of
diffractive reactions, by inclusion of specific, higher order Pomeron
exchange diagrams that account for single and double diffraction, and
the inclusion of hard processes, as described by QCD, that become more
important with energy.  Experiments at HERA have improved our knowledge
of the parton momentum distribution function inside a nucleon (see
e.g. ref. \cite{photonuclear}) which allow for a quantitative account of
the effect of hard interactions. This is vital when extrapolating
cross-sections and particle production up to the highest energies.  At
present only QGSJET, DPMJET and SIBYLL reach up to $> 10^{20}$ eV.

{\sc neXus}, the newest of the models, is constructed as a combined
effort of the authors of VENUS and QGSJET to model hadronic interaction
to the best of our knowledge, and to combine all necessary parts in a
consistent way. They start from the Universality hypothesis stating that
{\em the mechanisms of high-energy interactions are identical in
different type reactions}.  Thus one can study the final state parton
evolution and the subsequent hadronisation on the basis of $e^+e^-$
data, while the initial parton cascade can be tuned to deep-inelastic
lepton-proton data. In this way the hadronic interaction is broken into
separate building blocks which are deduced from simpler systems. This
constrains considerably the model parameters and ensures a more reliable
extrapolation.  {\sc neXus} is by construction a GRT-type model with
unified soft and hard interaction.  The former is described by the
traditional soft Pomeron exchange, the latter uses perturbative QCD
within the concept of the {\em semi-hard Pomeron}. In addition {\sc
neXus} employs a fully self-consistent treatment of energy and momentum
sharing between individual elementary scattering processes.

A model somewhere in between the purely phenomenological ones and the
GRT type models is SIBYLL \cite{sibyll}.  SIBYLL 1.6 was a so-called
{\em minijet} model, inspired by QCD, and treated the soft part rather
crudely.  It simulated a hadronic reaction as a combination of an
underlying soft collision, in which two strings are generated, and a
number of minijet pairs leading to additional colour strings with higher
transverse momentum ends.  In SIBYLL 1.6 the rise of the cross-section
with energy was solely due to the minijet production.  Recently the soft
part of SIBYLL was revised to allow for multiple strings from the soft
collision, leading to a contribution of soft interactions to the rise of
the cross-section, and for an energy dependent transverse momentum
cut-off for the minijet production \cite{sibyll2}.  With the new SIBYLL
2 a much better agreement of model predictions with experimental results
is achieved.

Cross-sections for nucleus-nucleus interactions are calculated on the
basis of the geometrical Glauber model \cite{glauber}. However, to
simulate particle production properly requires, for each
projectile nucleon, knowledge how often and with which of the target nucleons it
interacts. In more recent models this is determined via explicit
tracking of the nucleons in projectile and target during the collision.

The decay of colour strings into observable hadrons, as in particle
physics, is still a phenomenological procedure, but there is no reason
to assume that colour strings in cosmic ray showers should decay
differently from those from $e^+e^-$ or $p\bar{p}$ reactions. Therefore
those algorithms can be adopted that have been proven by particle
physics experiments.

Also the low-energy hadronic interactions are difficult to model.  While
for energies around 100 GeV many details are known about hadronic and
nuclear interactions, in the GeV range, where low particle
multiplicities dominate and resonances are important, only
phenomenological models exist.  Since most of the particles observed in
a CR experiment stem from interactions in the low-energy regime, these
models can influence predictions, too.  One program for low-energy
hadronic interactions is GHEISHA \cite{gheisha}, which was written
around 1985 to simulate the interactions of GeV secondaries from
$e^+e^-$ collisions with typical detector materials.  GHEISHA is a
phenomenological model, i.e. it was tuned to experimental results for a
variety of projectiles and targets in the few-GeV region and,
consequently, reproduces cross-sections and particle production rather
well. It is also used within the CERN detector simulation package GEANT
3 \cite{geant}.

\begin{figure}[b]
\begin{center}
\epsfig{file=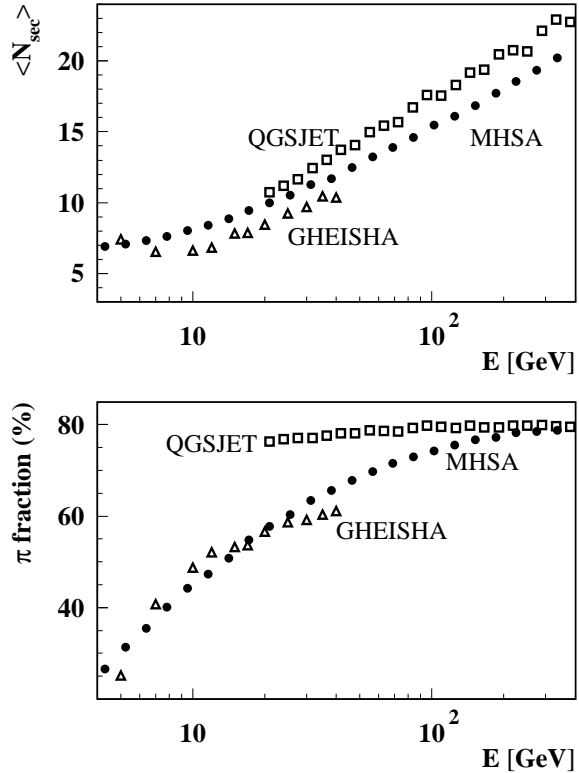, width=8cm}
\end{center}
\caption{Average number of secondaries and fraction of pions 
(charged + neutral) versus
primary laboratory energy in proton-air collisions. The triangles and
squares correspond to GHEISHA and QGSJET, respectively, and the full
circles show a (modified) HSA tuned to reproduce GHEISHA in the few-GeV
range and QGSJET far above 100 GeV.}
\label{fig-split}
\end{figure}
As an alternative program to GHEISHA for the low-energy hadronic
interactions the Ultra-relativistic Quantum Molecular Dynamics program
UrQMD \cite{urqmd} is considered which is developed and used for heavy
ion collision experiments. With it more detailed and theory-based
calculations can be performed, than with GHEISHA, in the few-GeV energy region.

A simpler, but very fast, approach is pursued by the Hillas Splitting
Algorithm (HSA) \cite{thin} in which the initial energy is split at
random into smaller and smaller portions. The secondary particles are
created from these energy packets, assigning their identity according to
externally provided probabilities.  In spite of its simplicity, the HSA
can be tuned to emulate very well some of the characteristics of the
secondaries emerging from hadronic collision, e.g. their multiplicity
and energy distributions. However, the HSA has to be complemented with
additional procedures to assign identity and transverse momentum to the
secondaries, and also the corresponding cross-sections must be provided
externally. The HSA can be adequately configured to reproduce the main
results from other, generally more involved, models.

Since GHEISHA was tuned in the few-GeV region and QGSJET uses the
Gribov-Regge approach, which works fully only in the range $E_{\rm
lab}>100$ GeV, the two models do not fit together smoothly. The
difference between the models ($\approx 25\%$) is usually well within
the fluctuations for a single interaction and the flexibility of HSA could
be used to {\em interpolate} between models for low and high energies in
the intermediate range where both are stretched to their limits.  This
is illustrated in Fig. \ref{fig-split}.  However, this example
demonstrates also that the HSA has no predictive power on its own and
needs a more advanced model according to which its parameters are
adjusted.

It holds in general, that the more complex a model is, the longer the
computing times become.  Especially in an air shower cascade, where many
millions of hadronic interactions occur, this may be prohibitive.
Therefore, two different approaches may coexist: reference-like models
that do the simulations as well as possible without consideration of
computing time, and faster but simpler models that are tuned in a first
step to reproduce the results of the reference.

\subsection{Inelastic cross-sections}

\begin{figure}[b]
\begin{center}
\epsfig{file=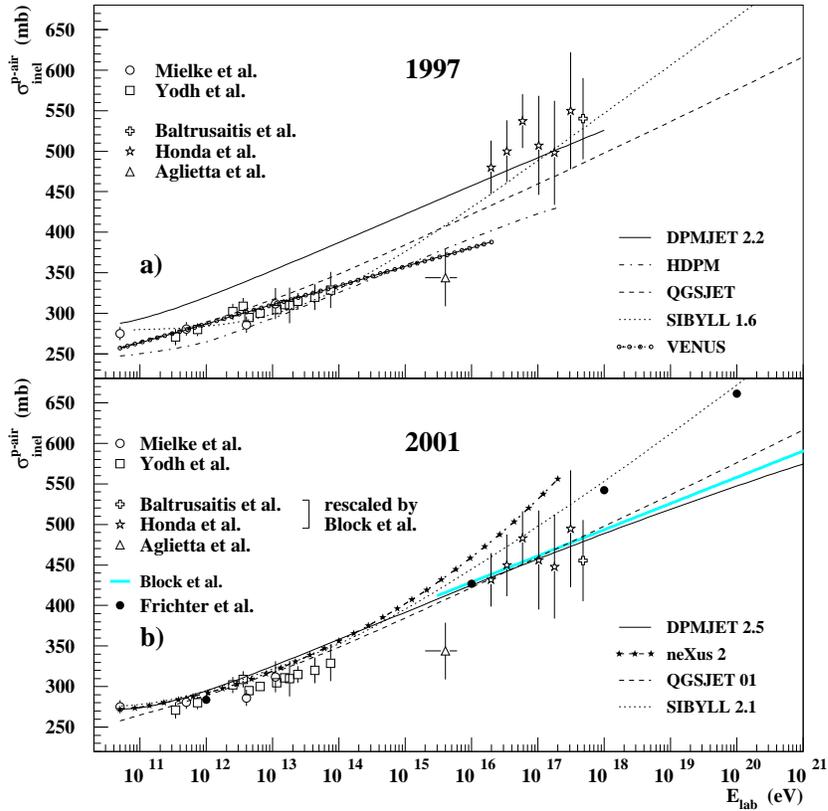,width=11cm}
\end{center}
\caption{\small Inelastic p-air cross-sections from measurements and
models. {\em Upper panel:} Situation in 1997. {\em Lower panel:} Situation
in 2001. Experimental data are taken from Refs.
\cite{mielke,yodh,baltrusaitis,honda,aglietta}, theoretical calculations
from Refs. \cite{block,frichter}.}
\label{fig-cross}
\end{figure}
A summary of experimental and predicted cross-sections is shown in Fig.
\ref{fig-cross}.  In 1997 the variation of the p-air cross-section in
experiments as well as in models amounted to about 25\% at $3\times
10^{15}$ eV, and to about 40\% at $10^{18}$ eV (shown in the upper
panel). For nucleus-air cross-sections the model uncertainty was only
10$-$15\% due to the averaging effect of many nucleons inside the
nucleus.  After some of the models have been revised new theoretical
predictions of cross-sections have been published. The situation is
shown in the lower panel of Fig. \ref{fig-cross}.  Some of the models
shown in the upper plot are no longer pursued by their respective
authors.

Recent calculations of the p-air cross-sections from Frichter et
al. \cite{frichter} agree well with the QGSJET cross-sections up to
about $10^{17}$ eV. Beyond this energy the predictions become
significantly higher than those of QGSJET, similar to the values of
SIBYLL.  Also Block et al. \cite{block} performed new calculations by
using a QCD inspired parameterization of accelerator data. They fitted
simultaneously the total p-p cross-sections, the ratio of real to
imaginary part of the forward scattering amplitude $\rho$, and the
nuclear slope parameter B. Then p-p cross-sections were converted via
Glauber theory into p-air cross-sections. Their extrapolation is shown
in Fig.  \ref{fig-cross}b and agrees very well with the cross-sections
used in the QGSJET model.

To extract cross-sections directly from EAS measurements is not
possible.  Usually the attenuation length $\Lambda$ of EAS in the
atmosphere is measured. Experiments select deeply penetrating showers to
retain only proton showers in their sample. The attenuation length
relates to the cross-sections $\sigma_{\rm inel}^{\rm p-air}$ via
$\Lambda = 14.6 \cdot {\rm k} \cdot {\rm m}_{\rm p} / \sigma_{\rm
inel}^{\rm p-air}$, where k depends strongly on the average inelasticity
and the inelasticity distribution of p and $\pi$ reactions and on the
energy.  While early and naive models fixed the inelasticity as one of
the free parameters, in GRT models the inelasticity distribution emerges
inevitably as a result of more fundamental properties of the interaction
model.  Since a value of k has to be assumed to derive the experimental
cross-sections, the final results are biased.  In addition, different
groups have been using different values for k. The experimental results
can be easily brought into agreement with each other and with the
theoretical calculations by slightly modifying k \cite{block}. For AGASA
and Fly's Eye data such a correction brought the experimental values in
the energy range $10^{16}-10^{18}$ eV down by 12$-$20\%, in good
agreement with the cross-sections as predicted by Block et al. or by
QGSJET.

As a result of various modifications the model predictions at knee
energies (i.e. $\approx 3\times 10^{15}$~eV) now agree to within about 6\%.

\subsection{Particle production}

The second important quantity that rules the shower development is the
inelasticity. This quantity combines the multiplicity and the energy
of the secondaries, thus describing how much of the energy of the incoming
particle is transferred onto secondary particles.  
Therefore it is more relevant than the particle multiplicity alone.
High inelasticity means that
the energy is dissipated quickly and the shower develops fast, i.e. it
reaches its maximum higher up in the atmosphere.  Low inelasticity means
that the leading particle
carries off most of the energy, 
leading to slow
developing and long showers. Obviously inelasticity and cross-section
influence the shower development in a very similar way, and are
therefore difficult to disentangle.
\begin{figure}
\begin{center}
\epsfig{file=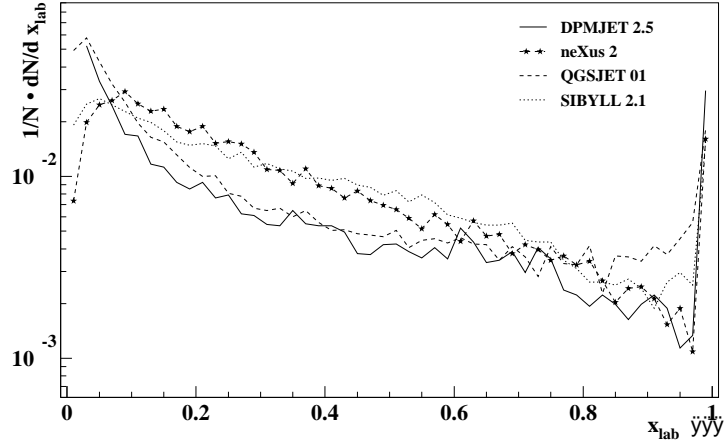,width=10cm}
\end{center}
\caption{\small Distributions of longitudinal momentum fraction carried
away by the leading baryon emerging from p-$^{14}$N collisions at
$E_{\rm lab} = 10^7$ GeV.}
\label{fig-inel}
\end{figure}
\begin{figure}
\begin{center}
\epsfig{file=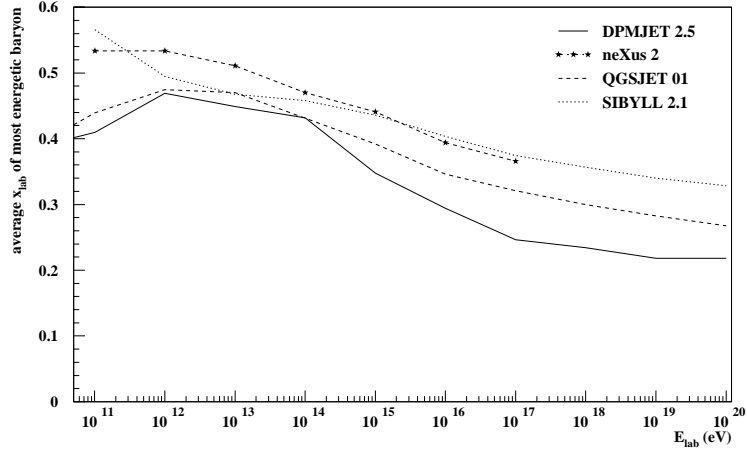,width=10cm}
\end{center}
\caption{\small Average of longitudinal momentum fraction carried away
by the leading baryon emerging from p-$^{14}$N collisions as function of
energy.}
\label{fig-avinel}
\end{figure}
In Fig. \ref{fig-inel} the distributions of the 
longitudinal momentum fraction, $x_{\rm
lab}$, that is carried away by the most energetic baryon in p-$^{14}$N
collisions are shown.  $x_{\rm lab}$ relates directly to the inelasticity
via $k_{\rm inel} = 1 - x_{\rm lab}$.  It is obvious from the figure that
the average value of $k_{\rm inel}$ alone does not fully describe the
distribution.  Indeed, the fluctuation in $k_{\rm inel}$ is one of the
major sources of shower fluctuations.  The distinctive peak near $x_{\rm
lab} = 1$ represents the diffractive events and reactions with large
$x_{\rm lab}$ are the ones that carry the energy efficiently deep into the
atmosphere. Since accelerator experiments barely see diffractive events
and their theoretical treatment allows some freedom, the models exhibit
a large spread in the range $x_{\rm lab} > 0.8$, which directly translates
into a systematic uncertainty in the shower analysis (see, for instance,
ref. \cite{raten}).  It is hoped that new experiments at LHC or RHIC
will measure particle production in the very forward direction and thus help
to reduce this uncertainty. The overall form of the elasticity
distributions is almost independent of the collision energy and their
average values change only slowly with energy.  All models show a
decrease (see Fig. \ref{fig-avinel}) from about 0.5 at $E_{\rm
lab} = 10^{12}$ eV to about 0.25 at 10$^{20}$ eV. The GRT-type
models predict lower values than SIBYLL.

\begin{figure}
\begin{center}
\epsfig{file=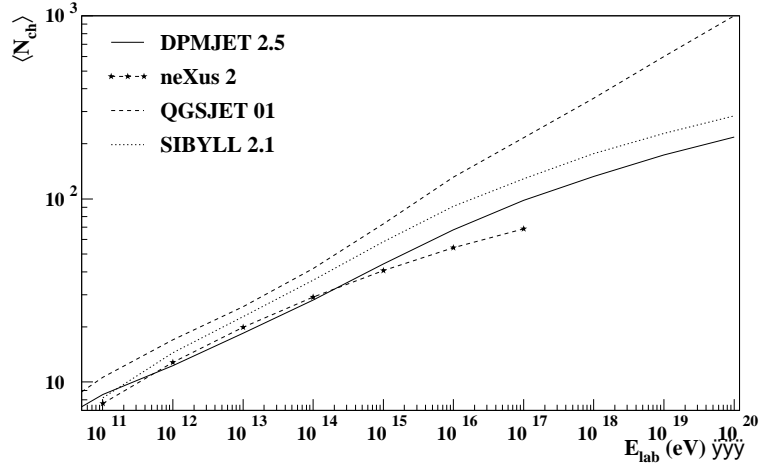,width=10cm}
\end{center}
\caption{\small Average charged multiplicity in $\pi$-$^{14}$N collisions 
as a function of energy as predicted by various models.}
\label{fig-mult}
\end{figure}
\begin{figure}
\begin{center}
\epsfig{file=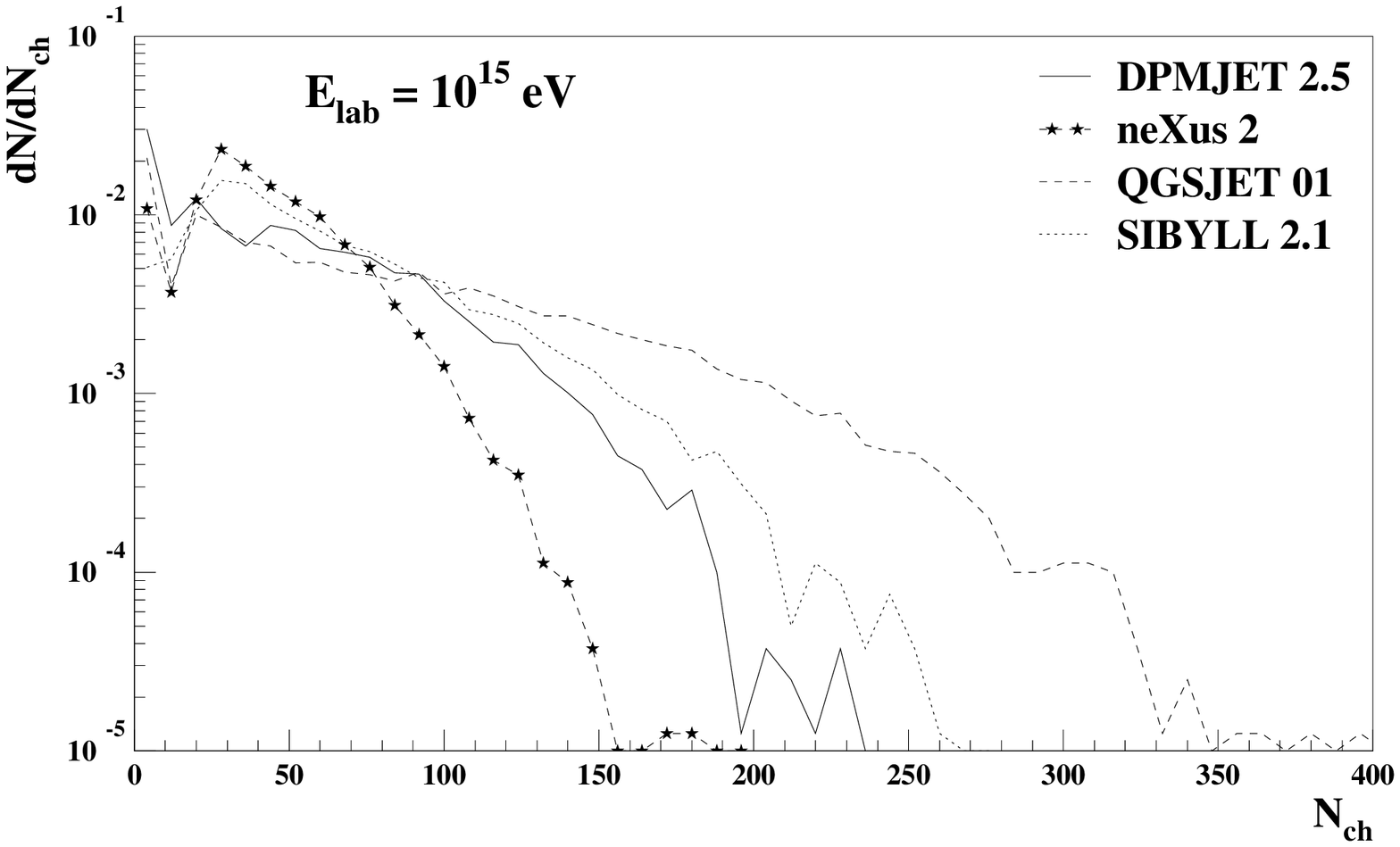,width=7.8cm} \hfill
\epsfig{file=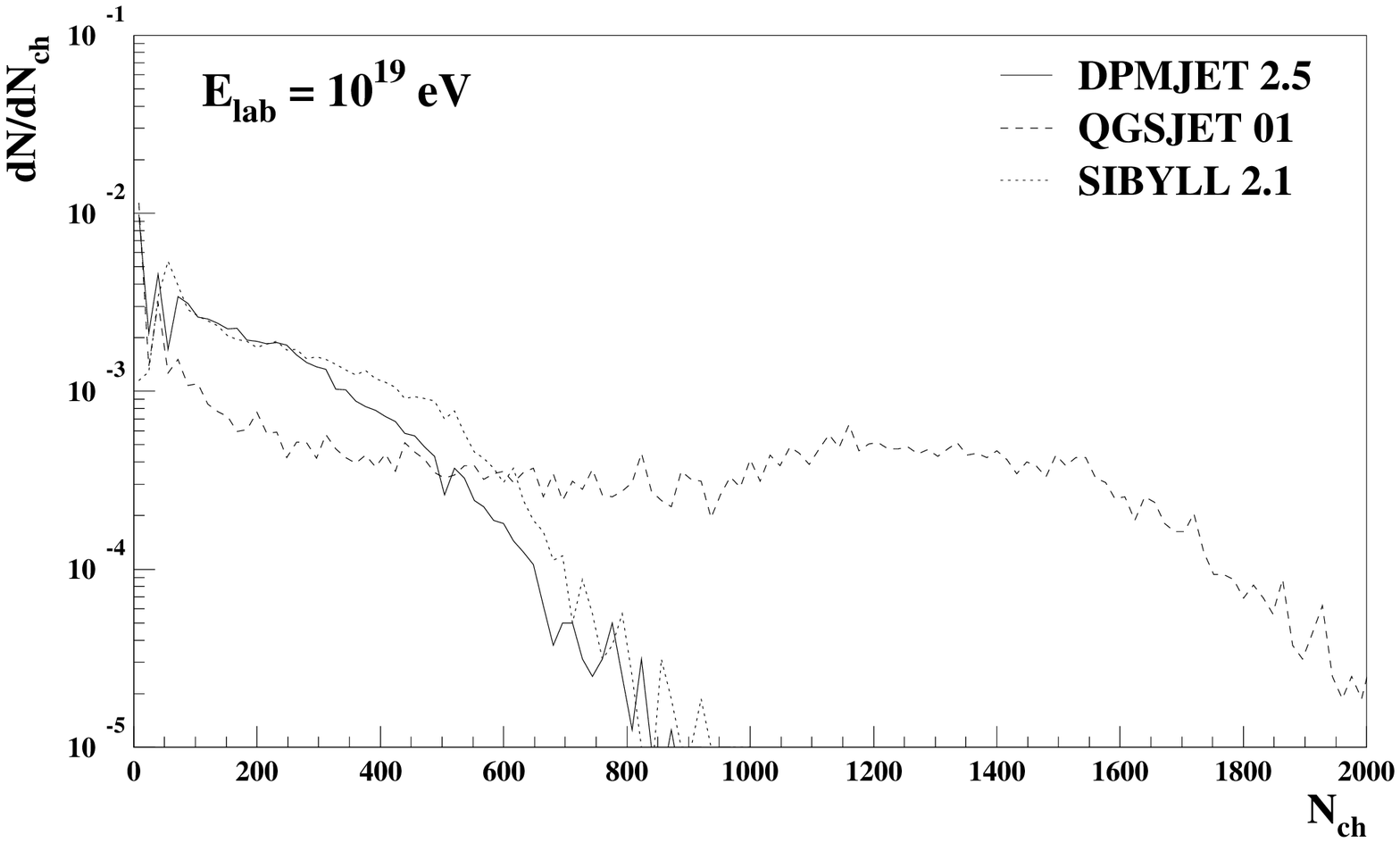,width=7.8cm}
\end{center}
\caption{\small Distribution of charged multiplicity of $\pi$-$^{14}$N
collisions at $10^{15}$  and $10^{19}$ eV as predicted by various models.}
\label{fig-multi19}
\end{figure}
In contrast to a widely-held belief the multiplicity of an hadronic
interaction on its own is not of great importance for the shower
development. Only the high-energy secondaries can influence the shower
development markedly, but secondaries of very low energies are irrelevant.
This can be demonstrated impressively by comparison of QGSJET with DPMJET
and SIBYLL. The average charged multiplicities of $\pi$-N collisions at
E$> 10^{19}$ eV differ by about a factor 5 (see Fig. \ref{fig-mult}) due 
to QGSJET interactions having a tail to very high numbers of secondaries 
which grows with energy. This is shown in Fig. \ref{fig-multi19} for 
$\pi$-N collisions at E$_{\pi} = 10^{15}$ and $10^{19}$ eV. Nevertheless the
fraction of energy converted into secondaries for QGSJET is in between the
values for SIBYLL and DPMJET (see Fig. \ref{fig-avinel}).
The reason is that most of the secondaries produced in QGSJET have low
energies and are produced in the central region rather than in the
forward region which is most important for the shower development. Also the
resulting longitudinal distributions of shower particles, as discussed in
the following section, show only relative small differences between QGSJET,
DPMJET and SIBYLL at $10^{19}$ eV (Fig. \ref{fig-long1e19}).

\subsection{Air shower development}

\begin{figure}[b]
\begin{center}
\epsfig{file=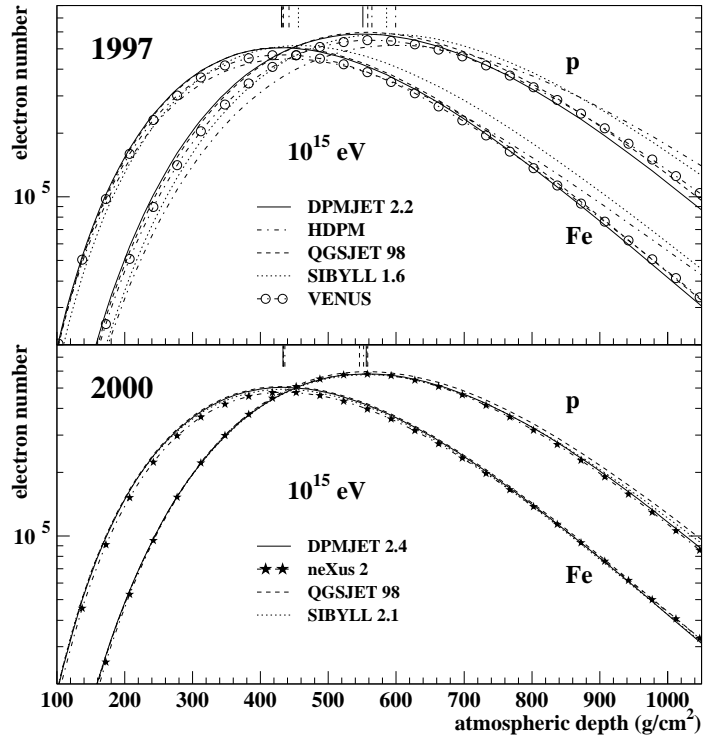,width=10cm}
\end{center}
\caption{\small Average longitudinal shower development of vertical
$10^{15}$ eV showers.  The positions of the shower maxima are indicated
by ticks at the upper rim of the picture.  {\em Upper panel:} Situation
in 1997. {\em Lower panel:} Situation in 2000.}
\label{fig-long}
\end{figure}

\begin{figure}
\begin{center}
\epsfig{file=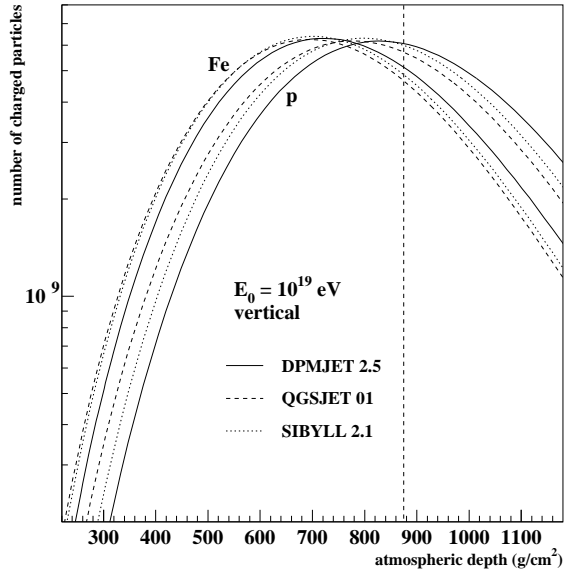,width=8cm}
\end{center}
\caption{\small Average longitudinal shower development for vertical
proton and iron showers of $10^{19}$ eV. The dashed vertical line
indicates the atmospheric depth of the Auger site.}
\label{fig-long1e19}
\end{figure}

The cross-section and the inelasticity determine the
longitudinal development of an EAS, which in turn is closely related to
the most important shower observables: the particle number at ground
level and their lateral distribution, the height of shower maximum 
${\rm X}_{\rm max}$, and the total
energy deposited in the electromagnetic component.  In Fig. \ref{fig-long}
average shower curves of 10$^{15}$ eV showers from different models are
shown.
The model convergence between 1997 and 2000 is clearly visible. While
for the 1997 models the position of ${\rm X}_{\rm max}$ varied by about
7.5\% for protons and 5.5\% for iron showers, with the new models the
variations are below 1\%. Also the particle numbers at ground level
agree much better than a few years ago.  The variations for p and Fe
showers were reduced from 80\% and 50\% to 15\% and 5\%, respectively.
Although $10^{15}$ eV is far below the energies relevant for the Auger
Observatory, Fig. \ref{fig-long} illustrates the spread where more
hadronic models are available.  A detailed comparison of SIBYLL 1.6 and
QGSJET 98 at energies $> 10^{19}$ eV has been reported in
ref. \cite{anchor}, where it is shown that the systematic uncertainty is
markedly larger. As illustration the shower curves of $10^{19}$ eV
proton and iron showers as obtained with the latest version of three
models are presented in Fig. \ref{fig-long1e19}.
The difference between the models is much larger, though it is still
smaller than the expected difference between proton and iron showers.
The SIBYLL 2.1 curve falls right between the predictions for two
GRT-type models.

\begin{figure}
\begin{center}
\epsfig{file=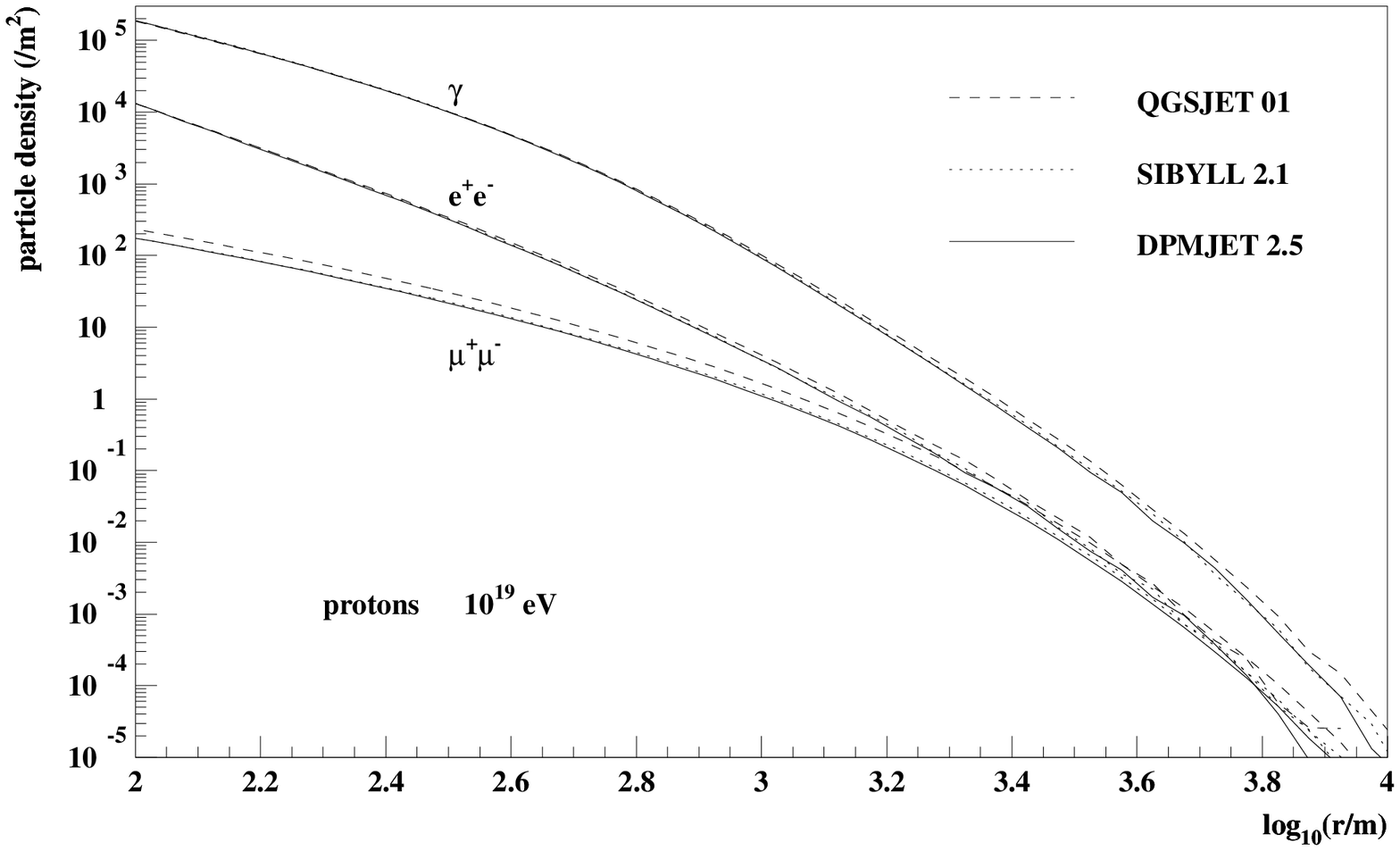,width=10cm}\\
\epsfig{file=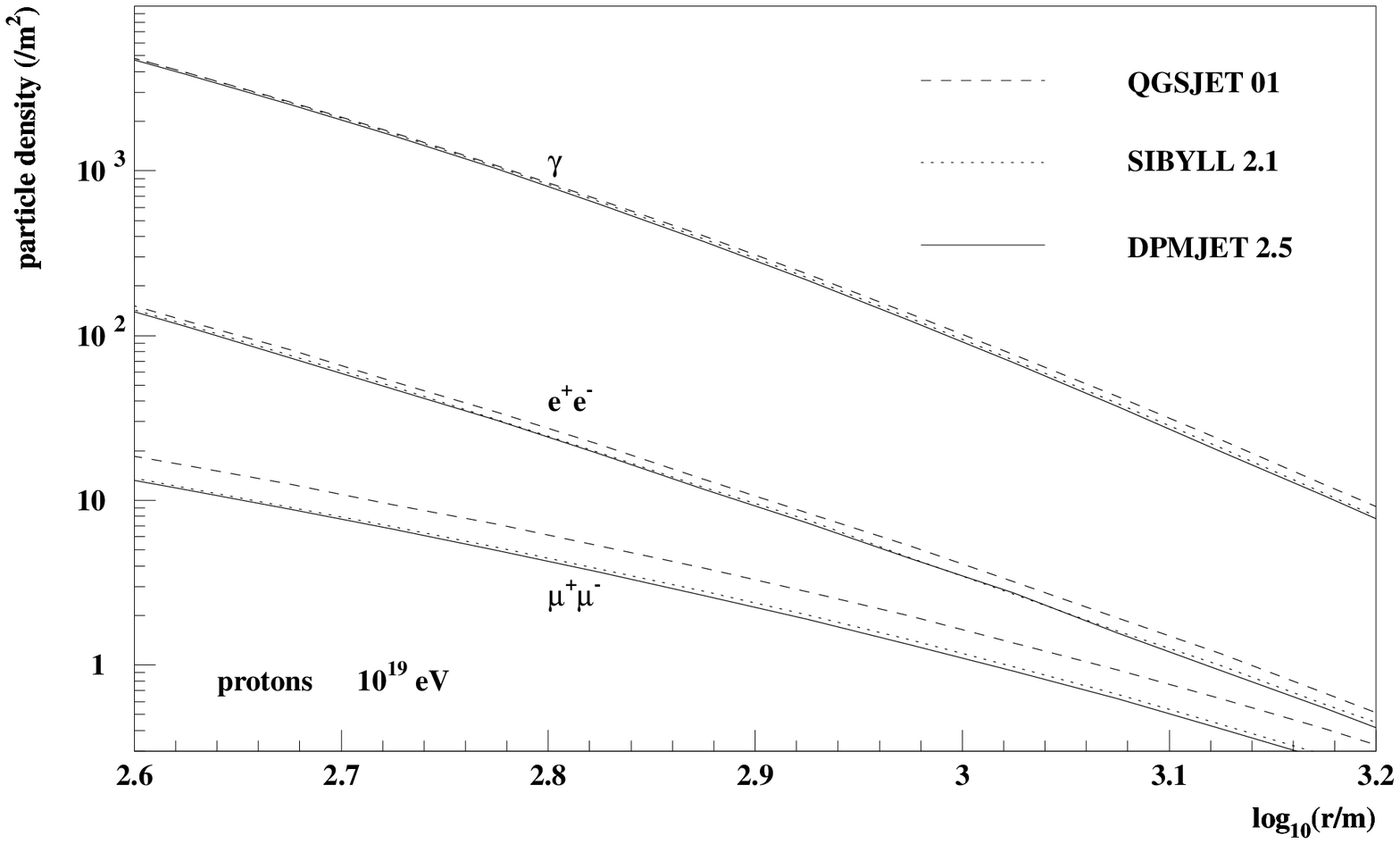,width=10cm}
\end{center}
\caption{\small Average lateral distributions of photons, electrons and muons
in vertical proton showers of $10^{19}$ eV. The lower panel shows the distance
range from 400 to 1600 m enlarged.}
\label{fig-latplot}
\end{figure}
Fig. \ref{fig-latplot} shows the lateral distributions of secondary photons,
electrons and muons in proton-induced showers of $10^{19}$ eV for different
models, in the radial distance range where Auger will record data.

DPMJET 2.5 and SIBYLL 2.1 distributions agree to better than 5\% for all
secondaries. QGSJET predicts systematically higher particle densities,
due to the long tail in the multiplicity distributions (see
Fig. \ref{fig-multi19}). 
Also the lateral distributions are slightly flatter 
(Fig. \ref{fig-latplot}). 
At a core distance of 1 km the photon and
electron densities differ by about 10$-$15\% and the muon densities by
about 30\%. For Auger this corresponds directly to an uncertainty in the
energy determination from the surface detectors of about 20\%. The
cross-calibration of surface detectors and fluorescence detector will
perhaps allow to discriminate models at this level of precision.

Also the models used for low-energy hadronic interactions introduce a
systematic uncertainty, e.g. with UrQMD there are about 30\% more
baryons in the energy range 2$-$10 GeV and factor of 2 less for 0.1$-$2 GeV
than with GHEISHA. Also UrQMD tends to produce 10\% more muons below 4
GeV and 10\% less above 4 GeV, the total number of muons stays
approximately the same in both cases.  As mentioned above, the
predictions of the HSA depend sensitively on the parameter setting,
e.g. the number of muons with $E_\mu < 10$ GeV can vary by up to 40\%.

\begin{figure}
\begin{center}
\epsfig{file=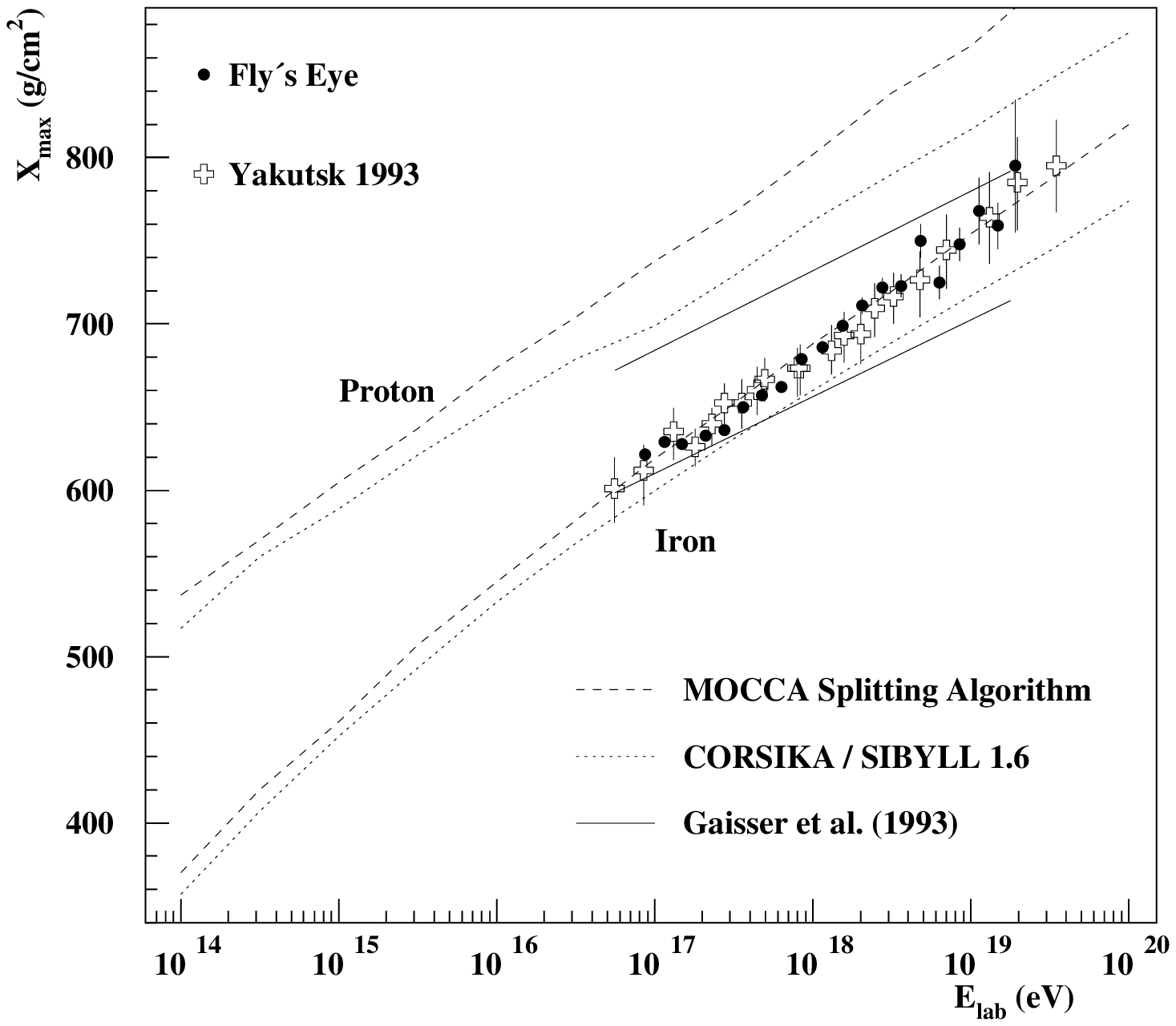,width=7.4cm} \hfill
\epsfig{file=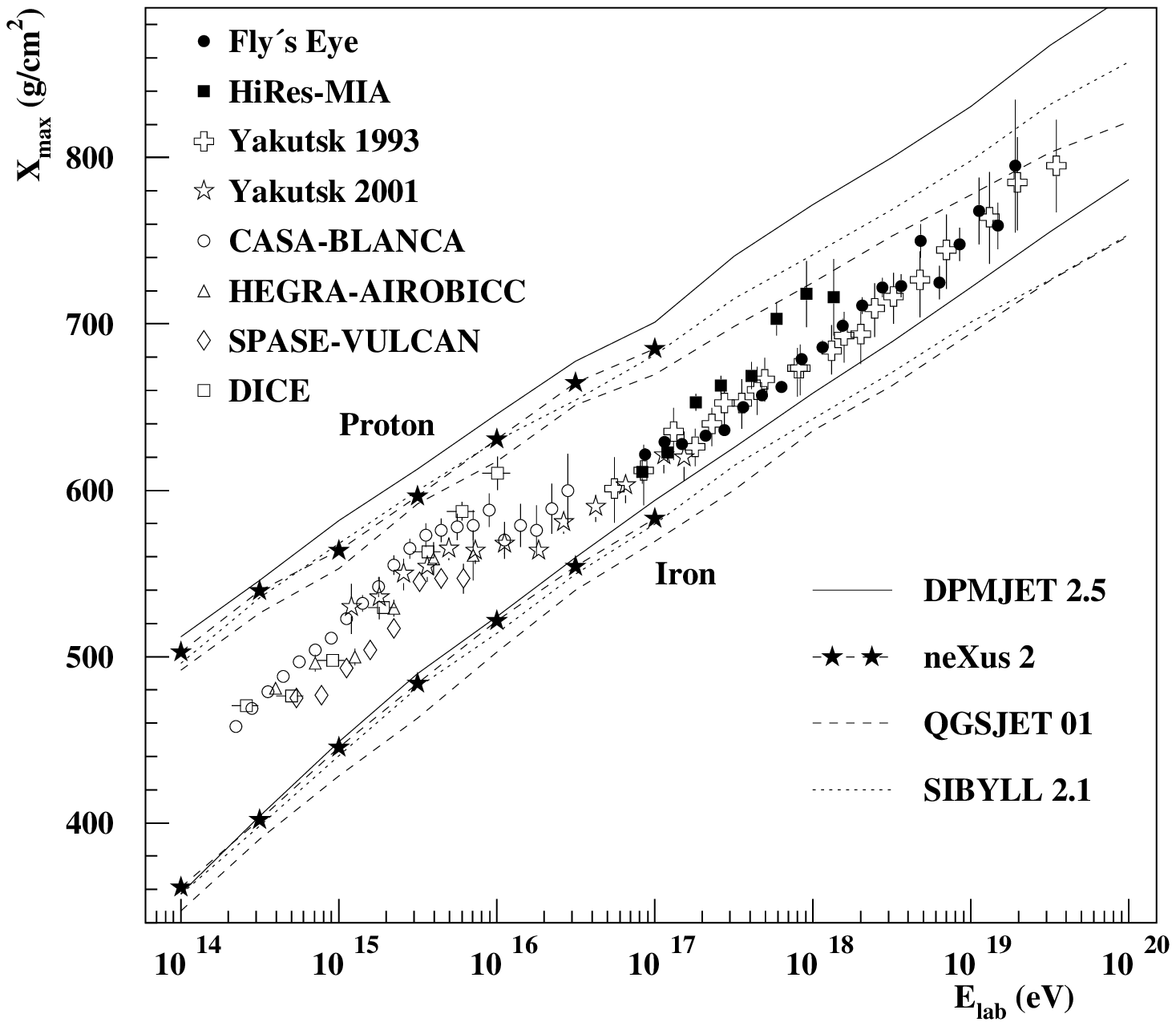,width=7.4cm}
\end{center}
\caption{\small Shower maximum ${\rm X}_{\rm max}$ as function of
energy.  {\em Left panel:} Measurements from Fly's Eye \cite{flys_eye}
and Yakutsk \cite{yakutsk}, compared with several models from the early
1990s \cite{mocca, pryke, gaisser}.  {\em Right panel:} measurements,
including newer data \cite{flys_eye,hires_mia,hires_mia2,yakutsk2,
casa_blanca,hegra,spase,dice}, compared with up-to-date models.}
\label{fig-xmax}
\end{figure}

The direct observation of the shower curve and the shower maximum is
possible with detectors that register the fluorescence or Cherenkov
light produced by the shower particles. These measurements are of great
advantage for the reconstruction of the shower energy and the primary
mass, but have a reduced duty cycle since they require clear and
moonless nights.  The Fly's Eye experiment pioneered this technique and
measured ${\rm X}_{\rm max}$ as a function of energy (see
Fig. \ref{fig-xmax}).  When first published  there were difficulties in
reproducing the experimental data with the available model predictions.
The Fly's Eye Collaboration interpreted their data by comparing to
simulations from Gaisser et al. \cite{gaisser}.  The MC predictions,
however, had to be shifted arbitrarily by 25 g/cm$^2$ to avoid particles
heavier than iron. Also the change of ${\rm X}_{\rm max}$ with energy,
i.e. the elongation rate,
was different in data and simulation. The different elongation rate
led to the conclusion that
the CR composition changed from basically pure Fe at $2\times 10^{17}$
eV to pure protons at $2\times 10^{19}$ eV (see left panel of
Fig. \ref{fig-xmax}).
When compared to MOCCA simulations using the HSA up to highest energies
the Fly's Eye data would have suggested a cosmic ray composition of iron
or even heavier. However,  it later became clear that the elongation rate, 
and therefore the interpretation of the data, was strongly model
dependent \cite{kalm,dawson}.

Meanwhile a number of experiments have provided data on ${\rm X}_{\rm
max}$ as a function of energy and the results are summarised in the
right panel of Fig. \ref{fig-xmax}.  More recent models can reproduce
the absolute values of ${\rm X}_{\rm max}$ over the whole energy range
and the change of mass composition seems much more moderate over the
Fly's Eye energy range.  However, a spread of the models of about 20 to
50 g/cm$^2$ for iron and proton showers, respectively, persists.

Close inspection of prediction of ${\rm X}_{\rm max}(E)$ from the modern
models (e.g. QGSJET 01 \cite{qgsjet,kalm,qgsheck}) reveals that the
superposition assumption, employed by some of the models to simulate
nuclear primaries, does not hold strictly. Especially for energies $>
10^{16}$ eV the elongation rate of iron induced showers is larger than
the one for proton showers ($\approx$ 62 and 56 g/cm$^2$ per decade,
respectively).  This is to be expected when the interactions of
individual nucleons in a nuclear collisions are treated correctly.  If
more than one projectile nucleon interacts with the same target nucleus,
all but the first nucleon will encounter a target that is excited above its
ground state.  The centre-of-mass
energy of the collision, and consequently the secondary particle
production, is smaller than in a superposition model where, by
definition, each projectile nucleon interacts with an undisturbed target
nucleus.

\subsection{Collider Results}

Another, and sometimes underestimated, source of uncertainties are the
errors of the collider results which are used to tune the interaction
models. There are a few instances where the uncertainty of these results
is revealed.

The inelastic p-$\bar{\rm p}$ cross-sections have been measured by three
different experiments at Fermilab \cite{cdf,e710,e811}.  Their values at
$\sqrt{s} = 1800$ GeV or ${\rm E}_{\rm lab}=1.7\times 10^{15}$ eV vary
from $80.03 \pm 2.24$ mb to $71.71 \pm 2.02$ mb.  This corresponds to
12\% uncertainty at an energy before the knee and clearly much more for the
highest-energy cosmic rays.

A second example is the pseudorapidity distribution of secondary
particles produced in p-$\bar{\rm p}$ collisions. The UA5 measurements
from $\sqrt{s}=200$ to 900 GeV \cite{ua5} have been widely used to tune
hadronic interaction models. GRT models always had severe difficulties
to reproduce the pseudorapidity densities dN/d$\eta$ near $\eta = 0$ and
the slope in dN/d$\eta$ at around $\eta =4$ at the same time.  However,
more recent measurements by Harr et al. \cite{harr} show a clearly
flatter distribution than the UA5 ones, that fits the GRT predictions 
very well \cite{knapp99}. In the near-forward region ($\eta \approx
4$) the differences in the measured particle densities amount to about
25\%.
 
Until new experimental data permit to clarify discrepancies like the
ones mentioned here, it seems rather unlikely that the basis of hadronic
models, and therefore their predictions, can become much better than
they are at present.

\section{Electromagnetic interactions}
\label{sec-em}

In contrast to the hadronic particle production, the electromagnetic
interactions of shower particles can be calculated very precisely from
Quantum Electrodynamics.  Therefore electromagnetic interactions are
not a major source of systematic errors in shower simulation.  Moreover
there exist very well tested packages to simulate these reactions in
great detail, such as EGS4 \cite{egs} or the electromagnetic parts of
GEANT \cite{geant}.  At the highest energies, however, a variety of effects
become important that are negligible at lower energies and are usually not
treated in the above mentioned programs.

High-energy photons can readily produce $\mu^+\mu^-$ pairs or hadrons.
The latter process may make a photon-induced shower look more like
hadronic ones.  The
muon pair production is calculated identically to $e^+e^-$ pair
production. The higher muon mass leads to a higher threshold for this
process.  QED firmly predicts the cross-section as function of energy
which approaches the Bethe-Heitler cross-section reduced by the factor
$(m_e/m_\mu)^2$.  The cross-section for hadron production by photons is
much less certain, since it involves the {\em hadronic structure} of the
photon.  It has been measured at the HERA storage ring for
photon energies corresponding to E$_{\gamma,{\rm lab}} = 2\times10^{13}$ eV 
\cite{ahmed,derrick}. This energy is still well
below Auger energies, but the experimental cross-sections constrain the
parameterisation used for extrapolation from the 100 GeV range.  At
these energies the cross-section for the process $\gamma \rightarrow
e^+e^-$ is $\approx 650$ mb, i.e. much larger than the cross-sections
for hadronic interaction ($\approx$ 1.4 mb) or for muon-pair production
($\approx 0.015$ mb).

Another effect to be considered in high-energy shower simulations is the
Landau Pomeranchuk Migdal (LPM) effect \cite{lpm}.  It describes an
interference effect between particle emission and scattering that occurs
only in matter. If the emission angle of the secondary particle is
smaller than a typical scattering angle, the secondary particle can be
readily re-absorbed. This leads to reduced cross-sections of pair
production by photons, and bremsstrahlung by electrons, at very high
energies, and to an effective prolongation of the radiation length,
which governs the mean free path of electrons and photons and determines
the development of electromagnetic showers.  The production of pairs of
secondaries of about equal energies is strongly suppressed.  In dense
media, like iron or lead, the LPM effect distorts electromagnetic
showers from about 10 TeV upwards.  However, since the upper atmosphere
is very thin the LPM effect becomes noticeable only for photons and
electrons of more than $10^{18}$ eV.  In hadronic showers the LPM
becomes important at about $10^{20}$ eV, since secondary photons
produced in the first interaction rarely have energies of more than 1\%
of the energy of the primary particle. Due to the LPM effect
electromagnetic sub-showers become longer and fluctuate more than normal
electromagnetic sub-showers.  The proper simulation of the LPM effect is
of special importance if one attempts to determine the elemental
composition from the longitudinal shower profile, or if one assumes that
the primary CR particles may be photons, as many of the scenarios that
have been constructed to explain the origin of highest-energy cosmic
rays predict.  For a recent discussion of the LPM effect in the context
of highest-energy air shower simulation see ref. \cite{lpm-aires}.

Muons of high energies initially lose energy through $e^+e^-$ pair
production and bremsstrahlung, at approximately equal rates.  The energy
loss is proportional to the muon energy and in air these processes
dominate the total muon energy loss only for $E_\mu > 20$ TeV. In
$10^{20}$ eV air showers the fraction of these high-energy muons is well
below 1\% of the total muon number. For a recent account of muon
bremsstrahlung and pair production at Auger energies see
ref. \cite{muon_brems_pair}.  High-energy muons also lose energy due to
hadronic reactions, though at a lesser rate than through bremsstrahlung
and pair production.  The hadronic energy loss is the least well-known
of all reactions of muons.  All these processes influence mostly the
muonic component of the shower.  They are therefore more important in
cases where the muon component is dominant, i.e. for very inclined
showers ($\theta > 70^\circ$) where very late stages of shower
development are observed, and where most of the electromagnetic and
hadronic shower particles have been absorbed.  For near-vertical
hadronic showers the above mentioned effects do not play a major role.

The spread of the shower particles at ground level is dominated by
multiple Coulomb scattering of charged particles off the nuclei in the
atmosphere. Air shower detectors for $10^{20}$~eV showers must be of
huge size and arrays of particle detectors are, for financial reasons,
built as sparse as the spread of shower particles allows. 
Consequently, the array detectors usually measure only particle
densities far away from the core (only 15\% of the shower cores fall
within distances $<300$ m from an Auger array detector). The particle
density at large core distances is determined by the tails of the
multiple scattering distribution.
Moli\`ere theory of multiple scattering  predicts realistic
distributions, with a roughly 
Gaussian distribution for small scattering angles, but larger tails from
single Coulomb scattering processes 
(for a short summary see Sec. 23 in ref. \cite{pdg}
and references therein).

As mentioned above, the measurement of fluorescence and Cherenkov light
in air, emitted by shower particles, allows a relatively reliable energy 
determination
of the primary particle. In the Auger experiment this is used for
the energy calibration of the  particle array with
fluorescence measurements.

For lower-energy showers it is possible to generate and track explicitly
the photons produced, but for the highest-energy cascades it is not
practical to propagate individual fluorescence or Cherenkov photons
since the number of normal shower particles already exceeds by far what
can be handled easily.  Therefore, during the shower simulation
distributions of shower particle location, energy and direction, as well
as energy deposits are recorded at various depths in the atmosphere
\cite{risseicrc}, which are used, in a second step, to simulate the
light production of an individual shower, the attenuation and scattering
in the atmosphere, and the image recorded at a fluorescence detector
site.

\section{Non-standard primaries and interactions}
\label{sec-neu}

The unknown origin of highest-energy cosmic rays has led to a variety of
more or less speculative explanations. They involve yet unknown decaying
superheavy  particles, known particles with non-standard properties, 
or even exotic new particles. Therefore,
shower simulations need to accommodate primaries other than the
classical protons and nuclei. Potential candidates are UHECR photons and
neutrinos, which may be produced in CR accelerators or as secondaries in
the decays of cosmic strings or superheavy relic particles
\cite{capelle}, neutrinos with hadron-like cross-sections, monopoles,
UHERONs or other exotic particles, for which cross-sections and reaction
processes have to be assumed.

If the highest-energy CRs indeed were photons then also reactions of the
primary photon with the Earth's magnetic field need to be considered.
Photons with $E > 10^{19}$ eV have a large probability to convert into
an $e^+e^-$ pair well before reaching the atmosphere \cite{mcbreen}.
The pair then strongly radiates in the field and produces a number of
lower-energy photons which again may convert into a pair. A pre-shower
is formed which will look markedly different in the atmosphere than a
shower from a single photon of the same energy. It will be wider, since
the pre-shower photons could spread out, and it will have a shorter
longitudinal development since the individual sub-showers are of lower
energy and do not suffer from LPM effect.  The formation of such
pre-showers, however, is usually not part of the shower simulation. They
are simulated with a suitable pre-processor that delivers the particles
which enter the atmosphere. These are then used as inputs of the
traditional air shower simulation programs \cite{bertou2}.

Because experiments like Auger survey such a huge volume of atmosphere
and because neutrino interaction cross-sections grow with energy, there
is sensitivity to showers induced by high-energy neutrinos
\cite{capelle,bertou}. The experimental signature would be normal
electromagnetic or hadronic showers with the exception that they start
very deep in the atmosphere. Especially for nearly horizontal showers,
where the atmosphere is $\approx$ 36000 g/cm$^2$ thick, no other primary
particle can penetrate the atmosphere to initiate a high energy shower
close to the detector and a selection of neutrino-induced showers seems
possible.  Also neutrino reactions are usually not part of shower
simulation programs.  Charged or neutral current neutrino-nucleon or
neutrino-electron interactions must be simulated externally and the
resulting particles can then be fed into the shower simulation. At
present some of these generators are prepared to investigate neutrino
induced showers in the Auger detector.

Additionally more exotic particles, such as magnetic monopoles or
supersymmetric particles, are discussed as primaries. In addition, the
first few interactions of conventional primaries may be energetic enough
to produce quark-gluon plasma or other exotic states with fundamentally
different number and energies of secondaries produced.  To simulate
these particles or reactions specific models must be constructed and
implemented in the shower simulation programs, e.g. by a pre-processor.

\section{Thinning}
\label{sec-thin}

\begin{figure}
\begin{center}
\epsfig{file=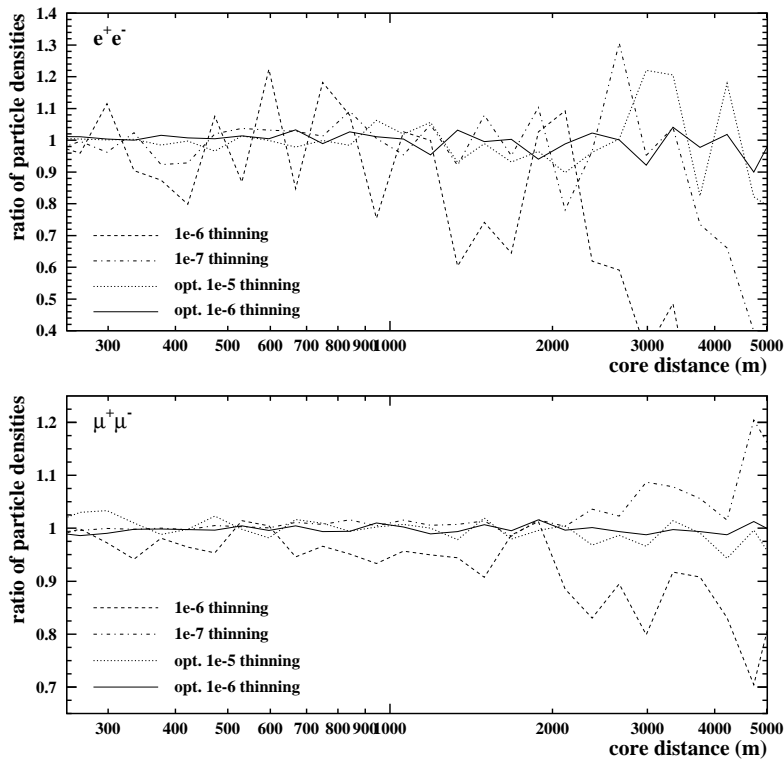,width=10.4cm}
\end{center}
\caption{Effect of the thinning energy on fluctuations of the lateral
distribution of $e^+e^-$ and $\mu^+\mu^-$.  Eight vertical proton
showers of $10^{19}$ eV have been averaged for each thinning level and
divided by a reference lateral distribution.  Results for
$\varepsilon_{\rm th}= 10^{-6}$ and $10^{-7}$ without weight limitation,
and for $\varepsilon_{\rm th}= 10^{-5}$ and $10^{-6}$ with optimum
weight limitation are shown.}
\label{fig-thin}
\end{figure}

For highest-energy air showers the number of secondaries becomes so
large ($> 10^{11}$) that it is prohibitive in computing time and disk
space to follow all of them explicitly and store the ones reaching
ground level. (To simulate a $10^{20}$ eV shower fully would take about
10 years per shower on a 750 MHz Linux workstation.)  Therefore, the
so-called {\em statistical thinning} was introduced by Hillas
\cite{thin}, which is a key concept in EAS simulations.  A small, but
representative, fraction of secondaries are fully tracked and all others
are discarded.  To account for the energy of the discarded particles
statistical weights are assigned to the tracked particles.  Thinning is
invoked whenever new particles are generated. If $E_A$ is the energy of
a particle producing secondaries of energy $E_{B_i}$, and $E_{\rm th}$ is
a fixed energy, called the thinning energy, then each particle in the
production vertex is selected for further tracking with a probability
$P_{i}$ as follows: If the energy $E_A$ of the primary particle is
greater than $E_{\rm th}$, then the probability $P_{i} = 1$ for $E_{B_i}
\ge E_{\rm th}$ and $P_{i} = E_{B_i}/E_{\rm th}$ otherwise. On the other
hand if $E_A$ is smaller than $E_{\rm th}$, meaning that the primary
particle comes from a previous thinning operation, then only one of the
secondaries is selected with probability $P_{i} = E_{B_i}/ \sum_{j=1}^n
E_{B_{j}}$. The weight of the accepted secondary particles is given by
the weight of the primary particle in the vertex divided by
$P_{i}$. This method ensures an unbiased sampling. Average values
calculated with the weighted particles do not depend on the thinning
energy, only fluctuations are artificially increased when thinning is
applied and depend on the thinning energy. The effect of the Hillas
thinning mechanism on the lateral distribution of electrons and
positrons from a $10^{19}$ eV vertical proton shower is shown in
Fig. \ref{fig-thin}.  The ratio of particle densities with thinning
levels $\varepsilon_{\rm th} = E_{\rm th}/E_{\rm prim} = 10^{-6}$ and
$10^{-7}$ to a reference density (representing densities without
thinning) are plotted.  The artificial fluctuations due to thinning,
showing up as fluctuations of the curve around 1, are clearly reduced
with the smaller thinning level.  The simulation done at
$\varepsilon_{\rm th} = 10^{-7}$ produced 10 times more output at ground
level and required $\approx$ 8.4 times more CPU time than that for
$\varepsilon_{\rm th} = 10^{-6}$.  The fluctuations are reduced by about
a factor 3$-$4 for the electromagnetic component.
For 10$^{19}$ eV showers with $10^{-7}$ thinning computing times still
range up to about 10 h per shower (on a 500 MHz machine) and secondary
particles at ground level carry weights of up to 10$^{7}$.
In its original form, the thinning algorithm
depends only on particle energies and the method is very effective close
to the shower core, where the particle densities are high.  With an
array of widely-separated surface detectors, however, most of the
detectors sample the air shower in a core distance range in which the
particle densities are rather small. Here thinning introduces large
artificial fluctuations. To reduce thinning fluctuations the
particle selection rule is modified to consider particle energies as
well as weights and the core distance.

A practical method for reducing the thinning fluctuations was
investigated by Kobal \cite{thin_kobal}. By limiting the weight
to a maximum value $w_{\rm lim}$ particles with higher weights can be
avoided. Particles with weights $\approx w_{\rm lim}$ are no longer thinned
and all their secondaries are tracked. The introduction of a weight
limit, while maintaining the thinning energy, increases the computing
time and the particle output again.  However, thinning energy and weight
limit can be optimised such that the sampling fluctuations become
minimal for a given computing time.  The optimum choice for most cases
is $w_{\rm lim} = E_{\rm prim} ({\rm in~GeV}) \cdot \varepsilon_{\rm
th}$ \cite{thin_kobal}.

For a $10^{19}$ eV proton shower, for example, the artificial
fluctuations for $\varepsilon_{\rm th} = 10^{-7}$ without weight
limitation are about 25\% larger than for $\varepsilon_{\rm th} =
10^{-5}$ with optimum weight limitation, and the computing time for the
latter is smaller by a factor 4.  To reduce the artificial fluctuations
further the thinning level can be reduced.  $\varepsilon_{\rm th} =
10^{-6}$ with optimum weight limitation needs twice as long as thinning
with $\varepsilon_{\rm th} = 10^{-7}$ without weight limitation, but the
artificial fluctuations are reduced by 75\%.  Thus, to optimise thinning
is of great advantage. The artificial fluctuations are about inversely
proportional to the square root of the number of particles recorded at
ground level.

The computing time is dominated by tracking electromagnetic particles
with average energies about 100 times lower than those of muons and
hadrons.  Consequently, electrons and photons carry weights that are
about 100 times higher than the muon and hadron weights. Therefore,
thinning and weight limitation can be applied differently to
electromagnetic and muonic or hadronic particles, e.g. limiting muons
and hadrons to a 100 times lower maximum weight and/or following them to
a smaller thinning energy.

An additional trick, the so-called radial thinning, reduces
the number of shower particles written to the
output file, while still performing the complete simulation.  
In the very dense region near the centre of the shower,
i.e. for $r< r_0$, the particle numbers are so large that 
detectors would be saturated for high-energy showers. The probability to
retain a particle is typically chosen to be $\propto (r/r_0)^k$ with
$r_0 \approx 150$ m and $k=2...6$. This method allows a reduction of
information stored by a factor between 2 and 5, 
without significant increase of
statistical fluctuations for the inner region, and with no loss at all at
large core distances, where usually most of the detectors are.

At present thinning seems unavoidable for coping with the huge number of
particles in a high-energy air shower. However, a severe disadvantage of
thinning becomes apparent when the shower particle lists are used to
calculate realistic detector responses, which is necessary to compare
simulations with experimental data.  It is non-trivial to simulate
detector signals and their spread in time from particles that carry 
large statistical weights,
and different methods are being developed to overcome this problem
\cite{billoir}.

\section{Characteristics of two simulation programs}
\label{sec-programs}

At present two program packages are available to simulate highest-energy
air showers. These are CORSIKA (COsmic Ray SImulation for KAscade)
\cite{corsika1,corsika2} and AIRES (AIR shower Extended Simulation)
\cite{aires}.  CORSIKA and AIRES both provide  fully 4-dimensional Monte
Carlo simulations of proton, photon, and nucleus-induced air shower
development in the atmosphere. Both simulate hadronic and
electromagnetic interactions, propagate particles through the curved
atmosphere, account for the Earth's magnetic field, for decays, energy
loss and deflection (and many less important processes), and produce
a list of all particles reaching ground level. Both programs assume the
same parametrisation of the US Standard Atmosphere as the atmospheric
model.

AIRES is originally based on MOCCA \cite{mocca}, but was rewritten and
significantly improved and extended.  The additions comprise links to
the external high-energy hadronic interaction models SIBYLL 1.6 and
QGSJET 98, the production of $e^+e^-$ pairs and bremsstrahlung by muons,
photonuclear reactions, the LPM effect for
high-energy $\gamma$ and $e^\pm$, and the simulation of exotic
primaries (e.g. $\nu$) \cite{bertou}.  For low-energy hadronic
interactions AIRES uses the Hillas Splitting Algorithm.

CORSIKA has been developed, over the last 12 years,  to become 
a standard analysis tool for
the air shower community. CORSIKA attempts to model the individual
processes of the shower development in as great detail as possible, to
some extent irrespective of the computing effort needed.  It employs
proven solutions wherever available. A variety of hadronic models have
been linked to CORSIKA and are used and updated to the specifications of
their respective authors.  The hadronic models HDPM \cite{cap,corsika1},
SIBYLL 2.1 \cite{sibyll2}, DPMJET 2.5 \cite{dpmjet25}, VENUS
\cite{venus}, QGSJET 01 \cite{qgsjet,kalm,qgsheck}, and {\sc neXus} 2
\cite{nexus,nexus1} are available above E$_{\rm lab} = 80$ GeV per
nucleon, and GHEISHA \cite{gheisha} and UrQMD \cite{urqmd} for energies
below.  CORSIKA uses an adapted version of EGS4 \cite{egs} for the
detailed simulation of electromagnetic interactions, that includes the
LPM effect, the production of muon pairs and hadrons by photons, muon
bremsstrahlung and $e^+e^-$ pair production by muons.  EGS was modified
to accommodate the variable atmospheric density, and to compute particle
production with double precision.  
The particles are followed to energies of typically 100 keV. In addition
the total energy deposited along the shower axis is recorded.
All two- and three-body decays, with
branching ratios down to 1\%, are modeled kinematically correct and
particle tracking and multiple scattering are done in great detail.  To
account for seasonal and geographical variations CORSIKA permits the
choice of a variety of atmospheric density profiles and the definition
of new ones.

AIRES uses its own procedures to simulate photo-electric effect, Compton
effect, electron bremsstrahlung, $e^+e^-$ pair production, and the
emission of knock-on electrons similar to what is done in EGS4 or GEANT.
Charged particle (multiple) scattering is treated by an effective
emulation of Moli\`ere's theory with finite nuclear size
corrections. The AIRES electromagnetic procedures work down to kinetic
energies of 85 keV. Particles below this threshold are discarded.

Both programs use a statistical thinning algorithm to keep computing
times and particle output at a manageable level. CORSIKA employs optimum
thinning following Hillas \cite{thin} and Kobal \cite{thin_kobal} while
AIRES uses its own approach \cite{aires} that also limits the
statistical weights.

AIRES, with the HSA and its own electromagnetic interaction routines, is
tuned to be  fast. AIRES is about of factor 3.5 faster than CORSIKA
when running with comparable (but non-optimum) thinning.  For
simulations of highest-energy showers with minimum thinning, computing
time may be the limiting factor and this difference in speed may prove
important. The particle output of AIRES is smaller than that of
CORSIKA. Both programs store 8 words of output information per particle
(i.e. particle id, p$_x$, p$_y$, p$_z$, x, y, t, weight). CORSIKA stores
each word with 32 bits (4 bytes), while AIRES provides the output in its
own reduced precision format with about 18 bits/word.  This may be of
advantage in case a large shower library is produced and the available
disk space is limited.

\section{Some Results}
\label{sec-results}

\begin{figure}[b]
\begin{center}
\epsfig{file=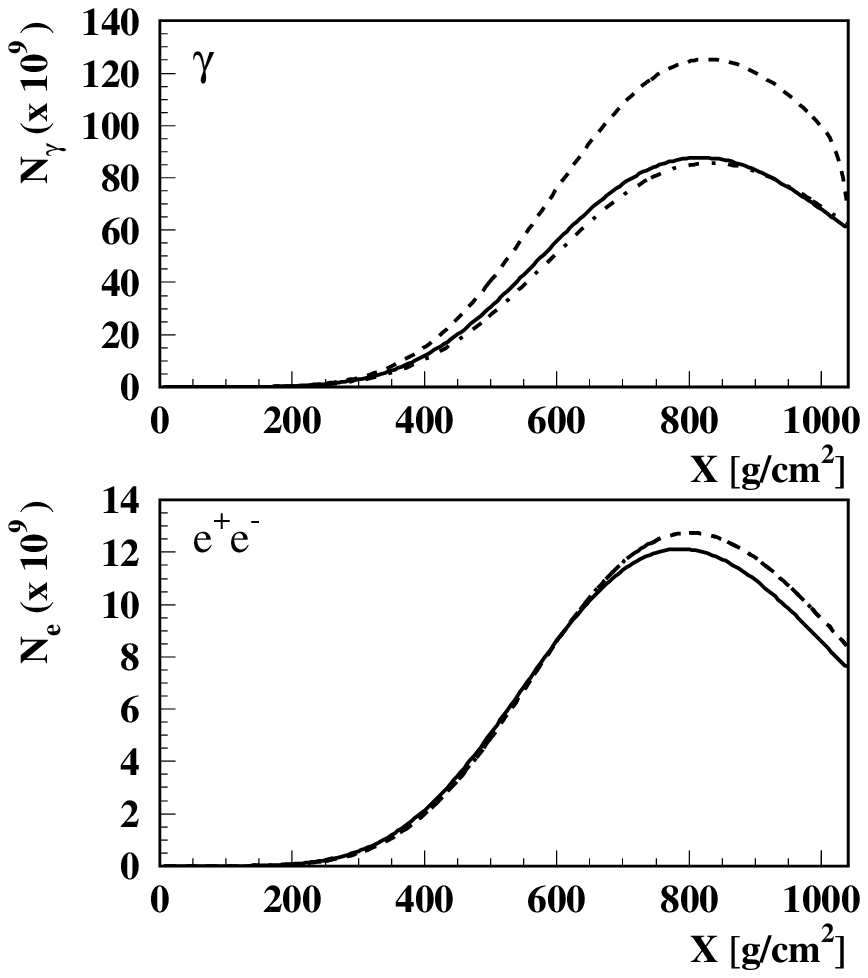,width=7cm}\\
\epsfig{file=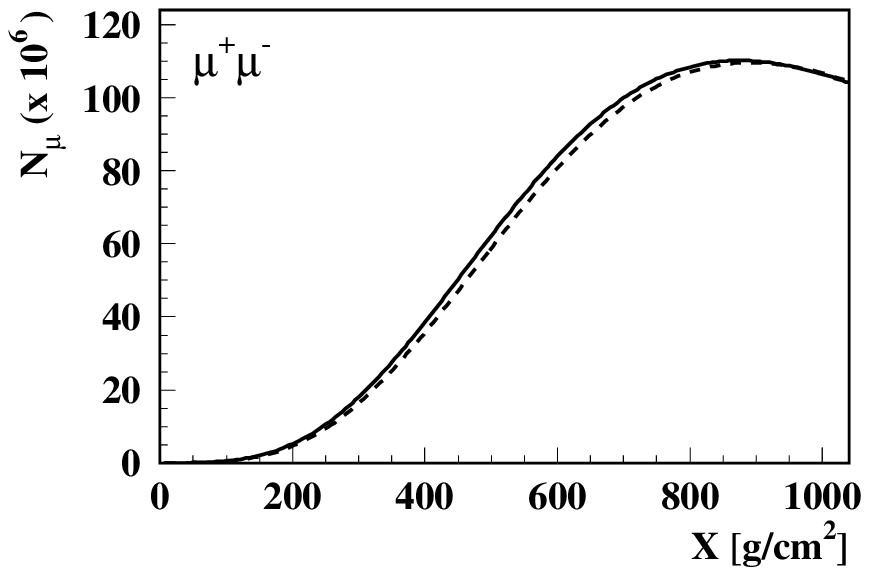,width=7cm}
\end{center}
\caption{Longitudinal development of N$_\gamma$, N$_{e^\pm}$
(E$_{\gamma,e^{+},e^{-}} \ge 0.1$ MeV) and N$_{\mu^\pm}$
(E$_{\mu^{+},\mu^{-}} \ge 100$ MeV) with atmospheric depth for vertical
proton showers of $2\times 10^{19}$ eV. CORSIKA: solid line, AIRES
including (excluding) upward going particles: dashed (dot-dashed) line.
For clarity the dot-dashed curves for $e^+e^-$ and $\mu^{+},\mu^{-}$ are
not shown. They agree with the solid line within 2\%.}
\label{fig-longe}
\end{figure}
\begin{figure}
\begin{center}
\epsfig{file=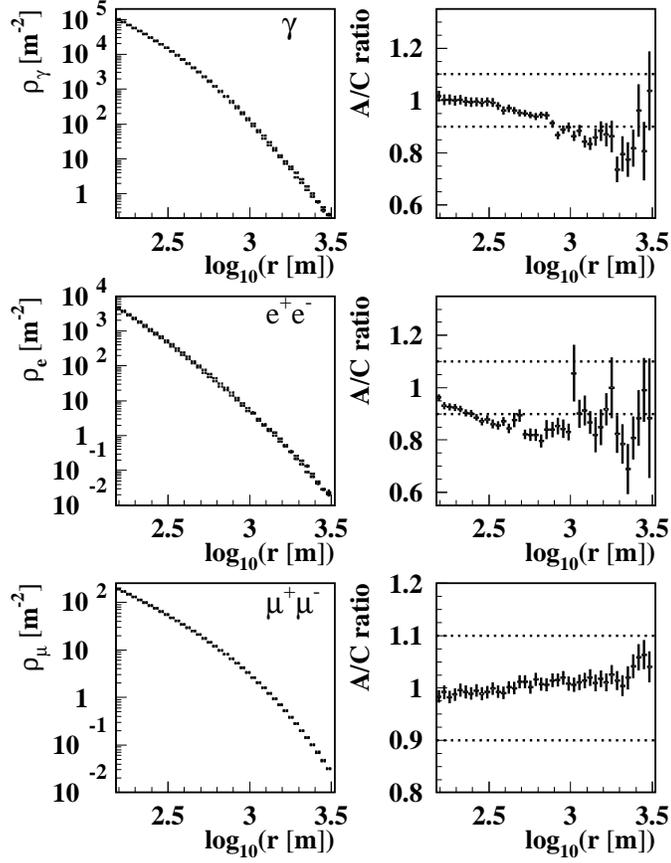,width=9cm}
\end{center}
\vspace*{-5mm}
\caption{Lateral particle densities for photons, electrons and muons
in vertical $2\times 10^{19}$ eV p showers.
{\em Left}: particle densities. 
The AIRES and CORSIKA points are virtually on top of each other.
{\em Right}: density ratios 
$\rho({\rm AIRES})/\rho({\rm CORSIKA})$ as function of core distance.}
\label{fig-lat1}
\end{figure}
\begin{figure}
\begin{center}
\epsfig{file=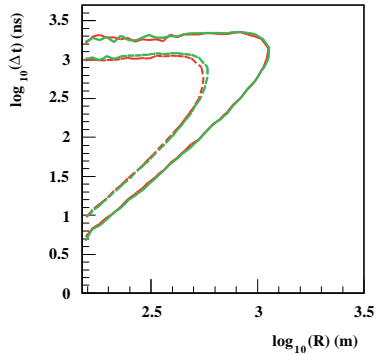,width=5cm}
\end{center}
\vspace*{-5mm}
\caption{Particle density contours for photons in the core
distance and arrival time plane, for vertical $2\times 10^{19}$ eV p showers.
CORSIKA: black, AIRES: grey.}
\label{fig-time}
\end{figure}
\begin{figure}
\begin{center}
\epsfig{file=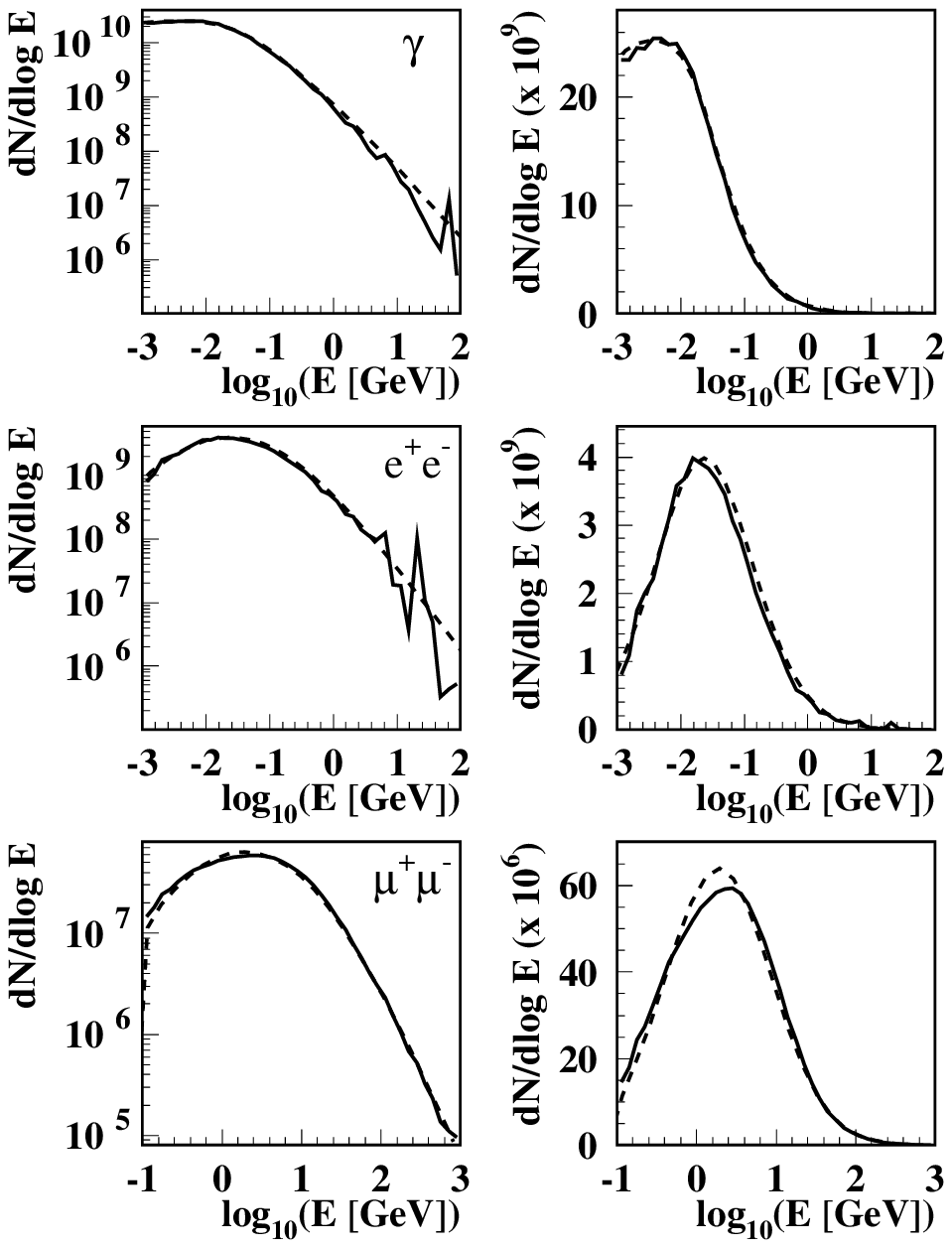,width=9.5cm}
\end{center}
\caption{Energy distributions for photons, electrons and muons.
{\em Left}: logarithmic abscissa, {\em Right}: linear abscissa.
CORSIKA: solid line, AIRES: dashed line.}
\label{fig-ener}
\end{figure}

Recently the performance  of CORSIKA and AIRES has been
compared for quantities that are measured in the AUGER experiment
\cite{aires-compare}. These are the longitudinal shower development, the
lateral distribution of particles far from the shower core, energy and
time distributions, all for different particle types.  Both programs
used the QGSJET 98 model for high-energy hadronic interactions ($E_{\rm
lab} > 80$ GeV). Averaged results for 100 vertical proton showers of 
$2\times 10^{19}$ eV with thinning at $\varepsilon_{\rm th} = 10^{-7}$ 
without weight limitation
are shown in Figs.  \ref{fig-longe},
\ref{fig-lat1} \ref{fig-time}, and \ref{fig-ener}.

\noindent
{\bf Longitudinal shower development.} The fluorescence light yield is
determined by the energy deposit in the atmosphere, which, in turn, is
dominated by the ionization due the numerous charged particles close to
the shower axis. Thus, the fluorescence light is closely related to the
total number of electrons (and positrons) as a function of depth. This
curve, however, is dominated by the high-energy model and how it
transfers the initial hadronic energy into the electromagnetic
channel. The longitudinal shower development is crucially dependent on
the inelastic cross-section and the inelasticity of interactions. Thus,
low-energy hadronic and electromagnetic models impose only second-order
effects on it. 

There is a large difference apparent in N$_\gamma$ as function of
depth. This is due to the fact, that in CORSIKA upward going particles
are discarded. These are predominantly very low-energy (sub MeV) photons
which contribute less than 2\% to the energy deposit in the
atmosphere. The disagreement in N$_\gamma$ vanishes almost completely if
AIRES discards the upward going particles (dot-dashed line).  At sea
level the photon number agrees to about 10\%.

The predicted evolution of N$_{e^+e^-}$ as a function of atmospheric
depth agrees well for the two programs, the electron numbers at the
maximum of the shower development differ by about 6\%, while the mean
positions of the maximum differ by about 25 g/cm$^2$ (see
Fig. \ref{fig-longe}).  This difference is also due to the different
treatment of upward going particles. If in both programs the upward
going particles are discarded the electron longitudinal distributions
agree within 2\%.

The muon numbers as function of depth, which sensitively depend on
details of the hadronic models, agree even better.  The differences at
the shower maximum are about 3\%.

\noindent
{\bf Lateral distributions.}  The Auger array detectors measure the
density of Cherenkov photons produced by shower particles in water. The
array detectors are positioned on a hexagonal grid with 1.5 km
separation. This means that array detectors will rarely be close to the
shower core.  Typically, particle densities will be recorded in the
range $r > 300$ m.  Fig. \ref{fig-lat1} shows the lateral densities of
secondary photons, electrons and muons. The density ratios, $\rho({\rm
AIRES})/\rho({\rm CORSIKA})$, on the right show the differences clearly.
With core distance CORSIKA tends to predict higher photon and electron
densities, reaching $\approx 20$\% for electrons at km distances. The
muon densities agree better.  At $r \approx 3$ km a deviation of about
3\% is observed.  Fig. \ref{fig-time} shows the photon distribution as
function of core distance and arrival time.  As expected, the larger the
core distance the later the particles arrive on average. The agreement
between AIRES and CORSIKA is very good.  The agreement between the
programs, despite the differences on the microscopic level, demonstrates
that particle densities at large core distances are mainly determined by
the transverse momenta at particle production and by multiple
scattering, and less by details of the low-energy models.

\noindent
{\bf Energy distributions.} 
The density of Cherenkov photons produced in a water tank by shower
particles (called the Cherenkov density) is approximately proportional
to the energy deposited and, therefore, depends not only on the particle
density but also on the energy the particles carry.  Electrons and
photons are basically absorbed in the water, i.e. deposit all their
energy (typically 1-100 MeV), while muons usually penetrate the tank and
release an energy of $\approx 2$ MeV/cm $\times$ their tracklength
(typically 240 MeV). Together with the fact that the muon density
decreases more slowly with $r$ than the electron and photon densities, this
means that the muon component is dominant at large core distances, as
measured by the tank response. Also the
energy distribution has a more direct relation to the low-energy
hadronic model than have longitudinal or lateral distributions, since the
form of the shower is basically determined from the higher-energy
interactions.  Fig. \ref{fig-ener} shows the energy distributions for
photons, electrons and muons in a logarithmic and a linear display. The
general agreement between AIRES and CORSIKA distributions is 
good. The most obvious discrepancies (in AIRES with respect to CORSIKA)
are a slight excess of photons and electrons with $E>10$ GeV, and a
deformation of the muon spectrum below 3 GeV, leading to a deficit for
muons with $E < 0.5$ GeV and an excess for $0.5< E < 3$ GeV. It is rather
likely that both discrepancies stem from the low-energy hadronic model, e.g
from the higher $\pi$ yield in the HSA as compared to GHEISHA.
        
The general agreement between AIRES and CORSIKA in longitudinal, lateral
and energy distributions is good. No discrepancies are found beyond the
20\% level.\\

\noindent
{\bf Models compared to experimental data.}  Though the comparison of
models between each other is an important exercise, an important test is
the comparison of simulations with experimental data. This requires a
detailed understanding of the detector response to the shower particles.
A variety of experiments have used modern shower and detector simulation
tools and have reported good overall agreement of model simulations with
measurements. In the following just three examples from different energy
regions will be shown.

\begin{figure}[b]
\begin{center}
\epsfig{file=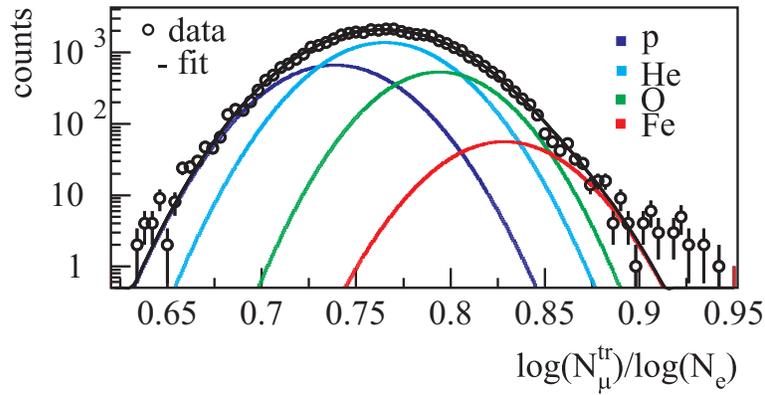,width=10cm}
\end{center}
\caption{Distribution of muon to electron ratio $\log_{10} N_\mu^{\rm
tr}/\log_{10} N_e$ in showers of energy $\approx 2\times 10^{15}$
eV. From ref. \cite{jenny_gauss}.}
\label{fig-jenny}
\end{figure}

The KASCADE experiment \cite{kascade} records air showers in the energy
range $10^{14} - 10^{16}$ eV and measures simultaneously the
electromagnetic, muonic and hadronic shower component with good accuracy.
This offers a good basis for a variety of model tests
\cite{kasc_modtest,raten}.  One of the quantities examined is the ratio
of muon to electron number.  Fig. \ref{fig-jenny} shows the measured
distribution of the muon to electron ratio $\log_{10} N_\mu^{\rm
tr}/\log_{10} N_e$ in showers with $E_0 \approx 2\times 10^{15}$ eV.
This distribution is compared with CORSIKA/QGSJET 98 simulations for p,
He, O and Fe primaries. The relative fractions of primaries are adjusted
to fit the experimental curve.  It is a great confirmation of the model
calculations that the muon-to-electron ratio of simulated protons
fits the left part, and the iron simulations fit, to some
extent, the right part of the experimental distribution, without any
arbitrary adjustment.
\begin{figure}
\begin{center}
\epsfig{file=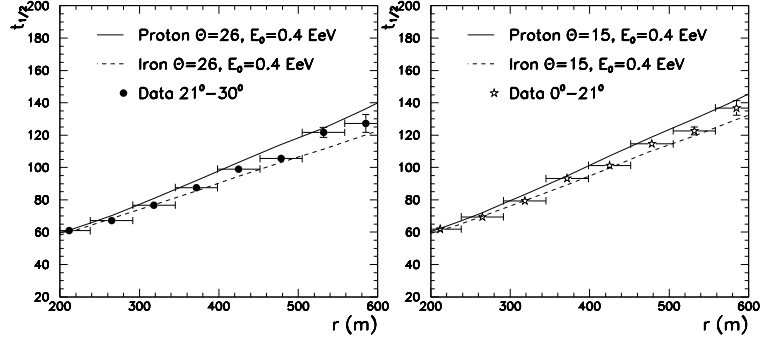,width=10cm}
\end{center}
\caption{Signal rise time as function of core distance for Haverah Park
data and CORSIKA/QGSJET 01 simulations for two different zenith
angles. From ref. \cite{ave_mass}.}
\label{fig-rise}
\end{figure}
\begin{figure}
\begin{center}
\epsfig{file=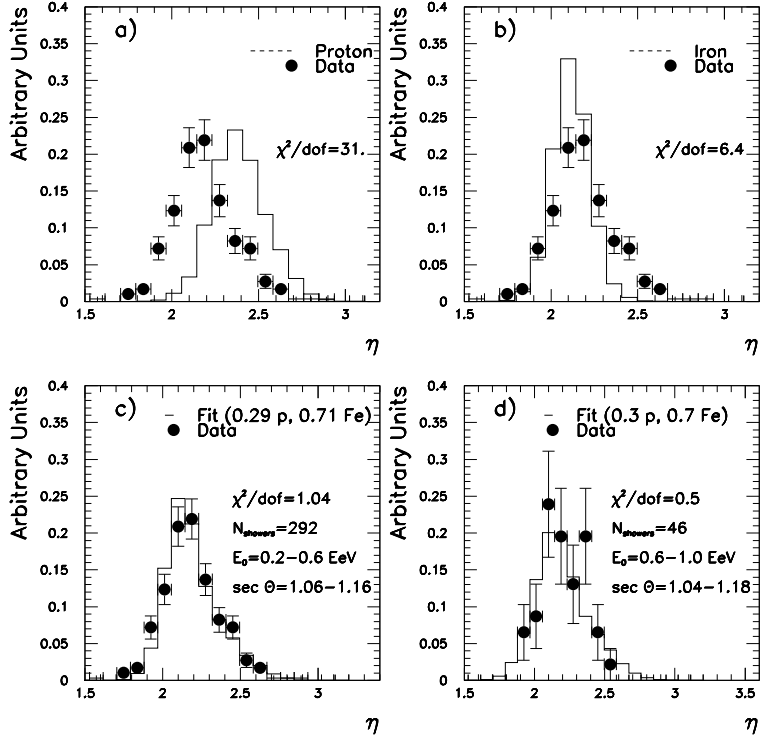,width=10cm}
\end{center}
\caption{Distribution of the shape parameter $\eta$ for Haverah Park
data and CORSIKA/QGSJET 01 simulations. The solid points represent
experimental data, the histograms the simulations.  a), b) and c) show
results for $E_0 = 0.2-0.6$ EeV, and d) for $E_0 = 0.6-1$ EeV.  From
ref. \cite{ave_mass}.}
\label{fig-eta}
\end{figure}

The Haverah Park experiment has measured air showers at energies between
$3\times 10^{17}$ eV and $10^{20}$ eV.  Composition studies in the 1980s
proved to be impossible since models at that time where not able to
reproduce the measured lateral distribution of shower particles at all.
Recently a sub-set of these data with $3\times 10^{17}$ eV $< E <
3\times 10^{18}$ eV has been re-analysed and interpreted by comparison
to CORSIKA/QGSJET 01 simulations \cite{ave_ener,ave_mass}.  Now
simulations reproduce the data very well.  Fig. \ref{fig-rise} shows the
signal rise times in water-Cherenkov detectors as a function of core
distance.  For the core distances shown the simulations for proton and
iron do not differ enough to allow a composition estimate, but the fact
that the measured data fall in between the simulations for protons and
iron indicates that the particle arrival times are well described by the
shower models.  Fig. \ref{fig-eta} shows the distribution of the shape
parameter $\eta$, which is used to parametrise the slope of the
water-Cherenkov signal as a function of core distance and is correlated
to the height of the shower maximum. The upper two panels show that the
data cannot be described by a pure proton or iron composition. For two
different energies the lower panels demonstrate that a mix of proton and
iron yields a good fit to the data.

Finally, even at energies up to $10^{20}$ eV the air shower models seem
to be close to the experimental data. A comparison of AGASA data with
CORSIKA/QGSJET 98 simulations \cite{nagano} shows that the form of the
average lateral distributions of muons and of all charged particles
agrees well for core distances from 300 to 2000 m.

\section{Summary}
In the last decade much has changed in air shower simulations. Through
better understanding of the relevant hadronic interactions and the
massive increase in computer performance very elaborate Monte Carlo
models became feasible that track individual particles through
atmosphere and detectors and simulate all their interactions
with matter in great detail.  Predictions of air shower programs have
become much more quantitative and describe the experimental data
sufficiently well for data analysis and the design of new experiments.

The weakest point in the simulations are the high-energy soft hadronic
interactions, which are not well examined at accelerators so far, and
the extrapolation to energies much beyond those available at
accelerators.  The most successful models are based on the Gribov-Regge
theory of multi-Pomeron exchange.  There is a clear trend of convergence
between different hadronic interaction models, due to objective
improvement of the physics input. However, a level of systematic
uncertainty of about 20\% is likely to persist, since many of the input
data, on which the models are built, appear to have uncertainties of
this size.

Once observed, larger inconsistencies between data and simulations offer a
possibility to improve the model assumptions.  The calorimetric
measurement of the shower energy via the fluorescence detectors in
coincidence with the charged particle measurements in the Auger
experiment will assist with stringent tests of the shower
models and promises the  determination of an absolute energy scale as well 
as setting constraints on the shower models.

Within the Auger Collaboration the two programs CORSIKA and AIRES are
used.  They account for all processes necessary for air showers
simulations up to 10$^{21}$ eV.  The programs complement each other in
the sense that CORSIKA uses more elaborate interaction models and AIRES
is faster. CORSIKA and AIRES agree to better than 20\% for the basic
shower parameters observed by the Pierre Auger Observatory.

\section*{Acknowledgements}
We are grateful for comments on an earlier version of this manuscript to
R. Engel, S. Ostapchenko, and A.A. Watson.
JK, DH and MR acknowledge support for a British-German Academic Research
Collaboration from The British Council and the DAAD.


\begin{thebibliography}{99}
\small
\bibitem{cdf} F. Abe et al., (CDF Collaboration), {\it Phys. Rev.} 
              {\bf D 50} (1994) 5550
\bibitem{hires_mia} T. Abu-Zayyad et al., {\it Phys. Rev. Lett.} 
              {\bf 84} (2000) 4276
\bibitem{hires_mia2} T. Abu-Zayyad et al., {\it  Astrop. J.} {\bf 557}
              (2001) 686
\bibitem{photonuclear} C. Adloff et al. (H1 Collaboration), 
              {\em Nucl. Phys.} {\bf B 497} (1997) 3
\bibitem{aglietta} M. Aglietta  et al., (EAS-TOP Collaboration),
              Proc.  $26^{th}$ Int. Cosmic Ray Conf., Salt Lake City, (USA)
              {\bf HE 1.3.04} (1999)
\bibitem{ahmed} T. Ahmed et al, (H1 Collaboration), {\it Phys. Lett.} 
             {\bf B 299} (1993) 374
\bibitem{ua5} G.J. Alner et al., (UA5 Collaboration), {\it Z. Phys.} 
              {\bf C 33} (1986) 1
\bibitem{e710} N.A. Amos et al., (E710 Collaboration), {\it Phys. Rev. 
              Lett.} {\bf 68} (1992) 2433
\bibitem{anchor} L.A. Anchordoqui et al., {\it Phys. Rev. D} {\bf 59}
              (1999) 094003
\bibitem{kasc_modtest} T. Antoni et al., (KASCADE Collaboration), 
              {\it J. Phys. G: Nucl. Part. Phys.} {\bf 25} (1999) 2161
\bibitem{raten} T. Antoni et al., (KASCADE Collaboration),
              {\it J. Phys. G: Nucl. Part. Phys.} {\bf 27} (2001) 1785 
\bibitem{hegra} F. Arqueros et al., (HEGRA Collaboration), {\it Astron. 
              Astrophys.} {\bf 359} (2000) 682
\bibitem{auger} Auger Collaboration, Auger Project Design Report, FNAL (1997)
\bibitem{ave_ener}M. Ave et al., astro-ph/0112253, {\it Astrop. Phys.}, in print
\bibitem{ave_mass}M. Ave et al., astro-ph/0203150, {\it Astrop. Phys.}, in print
\bibitem{e811} C. Avila et al., (E811 Collaboration), {\it Phys. Lett.} 
              {\bf B 445} (1999) 419
\bibitem{baltrusaitis} R.M. Baltrusaitis et al., {\it Phys. Rev.
              Lett.} {\bf 52} (1984) 1380
\bibitem{urqmd} S.A. Bass et al., {\it Prog. Part. Nucl. Phys.} 
              {\bf 41} (1998) 225;\\
              M. Bleicher et al., {\it J. Phys. G: Nucl. Part. Phys.} 
              {\bf 25} (1999) 1859
\bibitem{bertou} X. Bertou et al.,  {\it Astropart. Physics} {\bf 17}
              (2002) 183
\bibitem{bertou2} X. Bertou, P. Billoir, Pierr Auger technical note,
              {\bf GAP 1998-049} (1998)
\bibitem{billoir} P. Billoir, Reconstruction of Showers with the ground
              Array, Pierre Auger technical note {\bf GAP 2000-025} (2000)
\bibitem{flys_eye} D.J. Bird et al., {\it Phys. Rev. Lett.} {\bf 71} 
              (1993) 3401
\bibitem{block} M.M. Block, F. Halzen, T. Stanev, {\it Phys. Rev.}
              {\bf D 62} (2000) 077501
\bibitem{geant}  R. Brun et al., GEANT 3, Detector Description and Simulation
              Tool, CERN, Program Library CERN (1993)
\bibitem{cap} J.N. Capdevielle, {\it J. Phys. G: Nucl. Part. Phys.} 
              {\bf 15} (1989) 909
\bibitem{capelle} K.J. Capelle et al., {\it Astropart. Phys.} {\bf 8} (1998) 321
\bibitem{pdg} C. Caso et al., (Particle Data Group), {\it Review of 
              Particle Physics, Eur. Phys. J.} {\bf C3} (1998) 1
\bibitem{lpm-aires} A.N. Cillis, H. Franchiotti, C.A. Garcia Canal,
              S.J. Sciutto, {\it Phys. Rev.} {\bf D 59} (1999) 113012
\bibitem{muon_brems_pair} A.N. Cillis, S.J. Sciutto, {\it Phys. Rev.}
              {\bf D 64} (2001) 013010
\bibitem{dawson} B. Dawson et al., {\it Astrop. Phys.} 
              {\bf 9} (1998) 331
\bibitem{derrick} M. Derrick et al., (ZEUS Collaboration), {\it Phys. Lett.} 
              {\bf B 293} (1992) 465
\bibitem{spase} J.E. Dickinson et al., Proc. $26^{th}$ Int.
              Cosmic Ray Conf., Salt Lake City (USA), {\bf 3} (1999) 136
\bibitem{nexus} H.J. Drescher et al., {\it Phys. Rep.} {\bf 350} (2001) 93
\bibitem{yakutsk}M.N. Dyakonov et al., Proc. 23$^{rd}$ Int. Cosmic Ray Conf., 
              Calgary (Canada), {\bf 4} (1993) 303
\bibitem{sibyll2} R. Engel, T.K. Gaisser, T. Stanev, Proc. 26$^{th}$ Int.
              Cosmic Ray Conf., Salt Lake City (USA), {\bf 1} (1999) 415;\\
              R. Engel et al. (in preparation) 
\bibitem{gheisha} H. Fesefeldt, {\it The Simulation of Hadronic Showers 
              -Physics and Application-, Rheinisch-Westf\"alische 
              Technische Hochschule, Aachen} {\bf PITHA 85/02} (1985)
\bibitem{sibyll} R.S. Fletcher, T.K. Gaisser, P. Lipari, T. Stanev, 
              {\it Phys. Rev.} {\bf D 50} (1994) 5710;\\ 
              J. Engel, T.K. Gaisser, P. Lipari, T. Stanev, {\it Phys. 
              Rev.} {\bf D 46} (1992) 5013
\bibitem{casa_blanca} J.W. Fowler et al., {\it Astropart. Phys.} {\bf 15}
              (2001) 49
\bibitem{frichter} G.M. Frichter, T.K. Gaisser, T. Stanev, 
              {\it Phys. Rev.} {\bf D 50} (1997) 3135
\bibitem{gaisser} T.K. Gaisser et al., {\it Phys. Rev.} {\bf D 47} 
              (1993) 1919
\bibitem{glauber} R.J. Glauber, G. Matthiae, {\it Nucl. Phys.} {\bf B 21} 
              (1970) 135
\bibitem{gribov} L.V. Gribov et al., {\it Phys. Rep.} {\bf 100} (1983) 1
\bibitem{harr} R. Harr et al., {\it Phys. Lett.} {\bf B 401} (1997) 176
\bibitem{corsika1} D. Heck et al., {\it CORSIKA: A Monte Carlo Code to 
              Simulate Extensive Air Showers, Forschungszentrum Karlsruhe} 
              {\bf FZKA 6019} (1998)
\bibitem{qgsheck} D. Heck et al., (KASCADE Collaboration)
               Proc. 27$^{th}$ Int. Cosmic Ray Conf., Hamburg
              (Germany) (2001) {\bf HE 1.3} 233
\bibitem{thin} A.M. Hillas, Proc. 17$^{th}$ Int. Cosmic Ray Conf., Paris
              (France), {\bf 8} (1981) 193 \\
              Proc. 19$^{th}$ Int. Cosmic Ray Conf., La
              Jolla (USA), {\bf 1} (1985) 155 
\bibitem{mocca} A.M. Hillas, {\it Nucl. Phys. B (Proc. Suppl.)} {\bf 52B}
              (1997) 29
\bibitem{nexus1} M. Hladik et al., {\it Phys. Rev. Lett.} {\bf 86} (2001) 3506
\bibitem{honda} M. Honda et al., {\it Phys. Rev. Lett.} {\bf 70}
              (1993) 525 
\bibitem{qgsjet} N.N. Kalmykov, S.S. Ostapchenko, {\it Yad. Fiz.} 
              {\bf 56} (1993) 105;\\
              N.N. Kalmykov, S.S. Ostapchenko, {\it Phys. At. Nucl.} 
              {\bf 56} (3) (1993) 346;\\ 
              N.N. Kalmykov, S.S. Ostapchenko, A.I. Pavlov, {\it Bull. 
              Russ. Acad. Sci. (Physics)} {\bf 58} (1994) 1966 
\bibitem{kalm} N.N. Kalmykov, S.S. Ostapchenko, A.I. Pavlov,
              {\it Nucl. Phys. B (Proc. Suppl.)} {\bf 52B} (1997) 17 
\bibitem{kascade} H.O. Klages et al. (KASCADE Collaboration), {\it Nucl.
              Phys. B (Proc. Suppl.)} {\bf 52B} (1997) 92
\bibitem{knapp99} J. Knapp,  {\it Nucl. Phys. B (Proc. Suppl.)} 
              {\bf 75A} (1999) 89
\bibitem{corsika2} J. Knapp, D. Heck, {\it Extensive Air Shower Simulation 
              with CORSIKA: A User's Manual, Kernforschungszentrum Karlsruhe} 
              {\bf KfK 5196 B} (1993); for an up-to-date version see 
              http://www-ik3.fzk.de/\~{}heck/corsika/
\bibitem{yakutsk2} S. Knurenko et al., Proc. 27$^{th}$ Int. Cosmic Ray
              Conf., Hamburg (Germany) (2001) {\bf HE 1.3} 177
\bibitem{thin_kobal} M. Kobal, (Pierre Auger Collaboration), {\it Astropart. 
              Phys.} {\bf 15} (2001) 259
\bibitem{lpm} L.D. Landau and I.Ya. Pomeranchuk, {\it Dokl. Akad. 
              Nauk SSSR}  {\bf 92} (1953) 535 \& 735;\\
              A.B. Migdal, {\it Phys. Rev.} {\bf 103} (1956) 1811;\\
              E. Konishi et al., {\it J. Phys. G: Part. Phys.} 
              {\bf 17} (1991) 719
\bibitem{mcbreen}B. McBreen, C.J. Lambert, {\it Phys. Rev. D} {\bf 24}
              (1981) 2536
\bibitem{mielke} H.H. Mielke et al., {\it J. Phys. G: Nucl. Part.
              Phys.} {\bf 20} (1994) 637
\bibitem{nagano} M. Nagano et al., {\it Astropart. Phys.} {\bf 13} (2000) 277
\bibitem{egs} W.R. Nelson, H. Hirayama, D.W.O. Rogers, {\it The EGS4 
              Code System, Stanford Linear Accelerator Center} 
              {\bf SLAC 265} (1985)
\bibitem{pryke} C.L. Pryke, {\it Astropart. Phys.} {\bf 14} (2001) 319
\bibitem{dpmjet} J. Ranft, {\it Phys. Rev.} {\bf D 51} (1995) 64
\bibitem{dpmjet25} J. Ranft, preprints hep-ph/9911213 and hep-ph/9911232
              (1999)
\bibitem{risseicrc} M. Risse et al., Proc. $27^{th}$
              Int. Cosmic Ray Conf., Hamburg (Germany), (2001), 
              {\bf HE 1.5} 522
\bibitem{aires} S.J. Sciutto, preprint astro-ph/9911331 (1999);\\
              see also http://www.fisica.unlp.edu.ar/auger/aires
\bibitem{aires-compare} S.J. Sciutto, J. Knapp, D. Heck, Proc. $27^{th}$
              Int. Cosmic Ray Conf., Hamburg (Germany), (2001), 
              {\bf HE 1.5} 526
\bibitem{dice} S.P. Swordy, D.B. Kieda, {\it Astropart. Phys.} 
              {\bf 13} (2000) 137
\bibitem{jenny_gauss} J. Weber et al., (KASCADE Collaboration)
               Proc. $26^{\rm th}$ Int. Cosmic Ray Conf., Salt Lake City
               (USA) {\bf 1} (1999) 341
\bibitem{venus} K. Werner, {\it Phys. Rep.} {\bf 232} (1993) 87
\bibitem{yodh} G.B. Yodh et al., {\it Phys. Rev.} {\bf D 27},
              (1983) 1183
\end{thebibliography}
\end{document}